\documentclass[11pt,english]{article}
\usepackage[T1]{fontenc}
\usepackage[latin1]{inputenc}

\makeatletter

 \newcommand{\lyxaddress}[1]{
   \par {\raggedright #1 
   \vspace{1.4em}
   \noindent\par}
 }

\usepackage[margin=3cm,nohead]{geometry}
\usepackage{subfigure}
\usepackage{graphicx}

\usepackage{pst-node,lscape,amsfonts,amsbsy,graphics,lscape,mathrsfs,dsfont}

\newcommand{\e}{\mathrm{e}}

\renewcommand{\P}{\mathscr{P}}

\newcommand{\Tr}{\mathrm{Tr}}

\newcommand{\nn}{\nonumber \\}

\newcommand{\bea}{\begin{eqnarray}}
\newcommand{\eea}{\end{eqnarray}}
\newcommand{\bm}[1]{\mbox{\boldmath $#1$}}

\def\slash#1{\setbox0=\hbox{$#1$}  
   \dimen0=\wd0     
   \setbox1=\hbox{/} \dimen1=\wd1  
   \ifdim\dimen0>\dimen1   
      \rlap{\hbox to \dimen0{\hfil/\hfil}} 
      #1     
   \else     
      \rlap{\hbox to \dimen1{\hfil$#1$\hfil}} 
      /      
   \fi}      %

\usepackage{babel}

\usepackage{babel}
\makeatother
\begin{document}

\title{Transverse Quark Spin Effects and the Flavor Dependence of the Boer-Mulders
Function}

\author{Leonard P. Gamberg$^{(1)}$, Gary R. Goldstein$^{(2)}$, and Marc
Schlegel$^{(3)}$}

\maketitle

\lyxaddress{\begin{center}{\small $^{(1)}$ Division of Science, Penn State Berks,
Reading, PA 19610, USA}\\
 {\small $^{(2)}$ Department of Physics and Astronomy, Tufts University,
Medford, MA 02155, USA}\\
 {\small $^{(3)}$ Theory Center, Jefferson Lab, Newport News, VA
23608, USA}\end{center}}

\begin{abstract}
The naive time reversal odd (``T-odd'') parton distribution $h_{1}^{\perp}$,
the so-called Boer-Mulders function, for both up ($u$) and down ($d$) quarks
is considered in the diquark spectator model. 
While the results of different articles in the literature 
suggest that the signs of the Boer-Mulders function in semi-inclusive 
DIS for both flavors $u$ and $d$ are the same and negative, 
a previous calculation in the diquark-spectator model found 
that $h_{1}^{\perp(u)}$ and $h_{1}^{\perp(d)}$ have different signs. 
The flavor dependence is of significance for the analysis 
of the azimuthal $\cos(2\phi)$
asymmetries in unpolarized SIDIS and DY-processes, as well as for
the overall physical understanding of the distribution of transversely
polarized quarks in unpolarized nucleons. We find
substantial differences with previous work.
In particular we obtain  
half and first 
moments of  Boer-Mulders 
function that are negative over the full range in Bjorken $x$
for both the $u$- and $d$- quarks.
In conjunction with the Collins function we then predict the $\cos(2\phi)$
azimuthal asymmetry for $\pi^{+}$ and $\pi^{-}$ in this framework.
We also find that the Sivers $u$- and $d$- quark are negative and positive
respectively. As a by-product of the formalism,
 we calculate the chiral-odd but ``T-even''
function $h_{1L}^{\perp}$, which allows
us to present a prediction for the single spin asymmetry $A_{UL}^{\sin(2\phi)}$
for a longitudinally polarized target in SIDIS. 
\end{abstract}

\section{Introduction}

Naive time reversal-odd ({}``T-odd'') transverse momentum dependent
(TMD) parton distributions (PDFs) have gained considerable attention
in recent years. Theoretically 
it is expected that they  can account for  non-trivial transverse spin  and momentum correlations 
such as single spin asymmetries (SSA) in hard scattering processes
when transverse momentum scales  are on the order of 
intrinsic transverse momentum of quarks in hadron, namely 
$P_T\sim k_\perp  \ll \sqrt{Q^2}$.   Experiments are 
being performed~\cite{Airapetian:2004tw,Alexakhin:2005iw,Zhu:2006gx}
and proposed~\cite{Barone:2005pu,Afanasev:2007qh} 
to test these hypotheses by measuring 
transverse SSAs (TSSAs) and azimuthal asymmetries (AAs) 
in hard scattering
processes such as semi-inclusive DIS (SIDIS) or the Drell-Yan process (DY).
A prominent example of such a ``T-odd'' PDF is the Sivers
function $f_{1T}^{\perp}$ \cite{Sivers:1989cc,Sivers:1990fh} which
explains the observed SSA in SIDIS for a
transversely polarized proton target by the HERMES 
collaboration \cite{Airapetian:2004tw}. It correlates the intrinsic quark transverse momentum and the transverse nucleon spin.
The corresponding SSA on a deuteron target measured
by COMPASS \cite{Alexakhin:2005iw} vanishes, indicating a flavor dependence of the Sivers function.

Another  leading twist ``T-odd'' parton distribution, introduced
in~\cite{Boer:1997nt}, correlates the transverse spin of a quark
with its transverse momentum within the nucleon, 
the so called the Boer-Mulders function
$h_{1}^{\perp}$. It describes the distribution
of transversely polarized quarks in an unpolarized nucleon.  

Theoretically, twist two ``T-odd'' PDFs are of particular interest as
they formally emerge from the gauge link structure of the color gauge
invariant definition of the quark-gluon-quark correlation function~\cite{Collins:2002kn,Belitsky:2002sm,Boer:2003cm}.
This gauge link not only ensures a color gauge invariant definition
of correlation functions, but it also describes final state
interactions \cite{Brodsky:2002cx} and initial state interactions \cite{Boer:2002ju} which are necessary to generate SSA \cite{Brodsky:2002cx,Ji:2002aa,Boer:2003cm}.
Assuming factorization of leading twist SIDIS spin observables in
terms of the ``T-even''\cite{Mulders:1995dh} and ``T-odd'' TMD PDFs and
fragmentation functions (FFs), Ref. \cite{Boer:1997nt} shows how spin observables
in SIDIS can be expressed in terms of convolutions of these functions.
Formal proofs of factorization of leading
twist SIDIS spin observables were presented later in Refs. \cite{Ji:2004wu,Ji:2004xq,Collins:2004nx}.

Apart from the leading twist transverse SSA, measurements were also
performed in SIDIS on sub-leading SSA (i.e. they are suppressed like
$1/Q$, where $Q$ is the virtuality of the exchanged photon). In
particular, the asymmetry for a longitudinally polarized target was
measured by HERMES \cite{Airapetian:1999tv,Airapetian:2001eg,Airapetian:2002mf,Airapetian:2005jc},
whereas a non-vanishing beam-spin asymmetry was reported by 
CLAS \cite{Avakian:2003pk,Airapetian:2006rx}.
It was shown that final state interactions contribute also to these
types of single-spin asymmetries \cite{Afanasev:2003ze,Metz:2004je,Metz:2004ya,Afanasev:2006gw}.
Subsequently, this effect was described by the introduction of heretofore
unknown sub-leading twist ``T-odd'' PDFs \cite{Bacchetta:2004zf};
a complete list of these PDFs was presented in Ref. \cite{Goeke:2005hb}. 
These sub-leading twist ``T-odd'' PDFs discovered in this work were 
then incorporated into the tree-level formalism \cite{Bacchetta:2006tn}
completing the original work of \cite{Mulders:1995dh}.

In this paper we focus on the flavor dependence of the leading twist-2
{}``T-odd'' parton distributions in semi-inclusive DIS, i.e. Boer-Mulders function $h_{1}^{\perp}$, which is also
chirally odd, and the Sivers function $f_{1T}^{\perp}$ (keeping in mind that "T-odd" parton distributions in the Drell-Yan process flip their sign \cite{Collins:2002kn}). The Boer-Mulders function is particularly important for the analysis
of the azimuthal $\cos(2\phi)$ asymmetry in unpolarized SIDIS and
Drell-Yan. While in a partonic picture of the unpolarized $\cos(2\phi)$
asymmetry in SIDIS, the Boer-Mulders function is convoluted with the
``T-odd'' (and chiral-odd) Collins fragmentation function $H_{1}^{\perp}$
\cite{Collins:1992kk}, the corresponding $\cos(2\phi)$ asymmetry
in DY includes a convolution of the type $h_{1}^{\perp}\otimes\bar{h}_{1}^{\perp}$
\cite{Boer:1999mm} (where $\bar{h}_{1}^{\perp}$ is the Boer-Mulders
function for anti-quarks). Although these azimuthal asymmetries were
measured in SIDIS by the ZEUS collaboration \cite{Breitweg:2000qh,Chekanov:2006gt}
and in DY \cite{Falciano:1986wk,Guanziroli:1987rp,Conway:1989fs},
little is known about the Boer-Mulders function. Of particular interest
is the sign for different flavors $u$ and $d$ since this significantly
affects predictions for these asymmetries (see Ref. \cite{Lu:2006ew}).
The flavor dependence of $h_{1}^{\perp}$ was studied in the MIT-bag
model \cite{Yuan:2003wk} as well as in a spectator diquark model
\cite{Bacchetta:2003rz}, and a large $N_{c}$ analysis of TMDs was
performed in Ref. \cite{Pobylitsa:2003ty}. Model calculations of chirally odd generalized parton distributions (GPDs) \cite{Pasquini:2005dk} and a study of generalized form factors in lattice QCD \cite{Gockeler:2006zu} give indications about the flavor dependence of $h_{1}^{\perp}$ by means of non-rigorous and model-dependent relations between GPDs and transverse momentum dependent PDFs which were proposed and discussed in Refs. \cite{Burkardt:2005hp,Burkardt:2007xm,Meissner:2007rx}.
All of these theoretical and phenomenological treatments suggest an equal
(and negative) sign for the Boer-Mulders function for both $u$- and
$d$-quarks, with the exception of the calculation in the diquark
spectator model which results in opposite signs for $u$ and $d$.
The purpose of this paper is to consider the flavor dependence 
of $h_1^\perp$ and 
extend our earlier work on this subject~\cite{Goldstein:2002vv,Gamberg:2003ey}.
Additionally, we consider the flavor dependence of the ``T-even'' function, $h_{1L}^\perp$ 
which is also of interest in exploring the transverse momentum and quark spin correlations
in a longitudinally polarized target~\cite{Kotzinian:1997wt}.

\section{{}``T-Odd'' PDFs in the Spectator Model}

Transverse momentum quark distribution and fragmentation functions
contain essential non-perturbative information about the partonic
structure of hadrons. Practically speaking their moments are calculable
from first principles in lattice QCD. A great deal of understanding
has also been gained from model calculations using the spectator framework.
In addition to exploring the kinematics and pole structure of the
TMDs~\cite{Metz:2002iz,Gamberg:2003eg,Collins:2004nx} phenomenological
estimates for parton distributions \cite{Jakob:1997wg} and fragmentation
functions \cite{Ali:1995vw} for ``T-even'' PDFs and for ``T-odd'' PDFs \cite{Brodsky:2002cx,Ji:2002aa,Goldstein:2002vv,Gamberg:2003ey,Boer:2002ju,Bacchetta:2003rz,Gamberg:2006ru,Afanasev:2006gw,Meissner:2007rx}
have been performed. We extend these studies to explore the flavor dependence
of the ``T-odd'' pdfs adopting the factorized approach used in Refs. \cite{Ali:1995vw,Jakob:1997wg,Gamberg:2003ey}.

We start (cf. \cite{Jakob:1997wg}) from the definition of the 
fully-unintegrated, color gauge invariant, quark-quark correlator\[
\]
\begin{equation}
\Phi_{ij}(p;P,S)=\sum_{X}\int\frac{d^{4}\xi}{(2\pi)^{4}}\;\e^{ip\cdot\xi}\langle P,S|\,\bar{\psi}_{j}(0)\,\mathcal{W}[0\,|\,\infty,0,\vec{0}_{T}]\,|X\rangle\langle X|\,\mathcal{W}[\infty,\xi^{+},\vec{\xi}_{T}\,|\,\xi]\,\psi_{i}(\xi)\,|P,S\rangle,\label{eq:Fully-unQQ}\end{equation}
 where the gauge link indicated by the (straight) Wilson
line is given by
\begin{equation}
\mathcal{W}[a\,|\, b]=\P\exp\left\{ -ig\int_{a}^{b}ds^{\mu}\; A_{\mu}(s)\right\} .\end{equation}
 In an arbitrary gauge there is a Wilson line at light cone infinity
pointing in transverse directions \cite{Belitsky:2002sm,Boer:2003cm}.
Here, we work in Feynman gauge where the transverse Wilson line
vanishes \cite{Belitsky:2002sm}. In the definition (\ref{eq:Fully-unQQ})
we insert a complete set of intermediate states $\mathds{1}=\sum_{x}|X\rangle\langle X|$.
In the diquark model the sum over a complete set of intermediate on-shell
states $|X\rangle$ is represented by a single one-particle diquark
state $|\, dq;\, p_{dq},\lambda\rangle$, where $p_{dq}$ is the diquark
momentum and $\lambda$ its polarization. Since the diquark is {}``built''
from two valence quark it can be a spin 0 particle (scalar diquark)
or a spin 1 particle (axial-vector diquark). By applying a translation
on the second matrix element in Eq. (\ref{eq:Fully-unQQ}) we can
integrate out $\xi$, perform the momentum integration over the diquark
momentum $p_{dq}$, and obtain \begin{eqnarray}
\Phi_{ij}(p;P,S) & = & \sum_{\lambda}\frac{\delta((P-p)^{2}-m_{s}^{2})\Theta(P^{0}-p^{0})}{(2\pi)^{3}}\,\langle P,S|\,\bar{\psi}_{j}(0)\,\mathcal{W}[0\,|\,\infty,0,\vec{0}_{T}]\,|\, dq;\, P-p,\lambda\rangle\times\nonumber \\
 &  & \langle\, dq;\, P-p,\lambda|\,\mathcal{W}[\infty,0,\vec{0}_{T}\,|\,0]\,\psi_{i}(0)\,|P,S\rangle.\label{eq:Correlator}\end{eqnarray}
 The essence of the diquark spectator model is to calculate the matrix
elements in Eq. (\ref{eq:Correlator}) by the introduction of effective
nucleon-diquark-quark vertices.

\begin{figure}
\begin{center}
\includegraphics[scale=0.4]{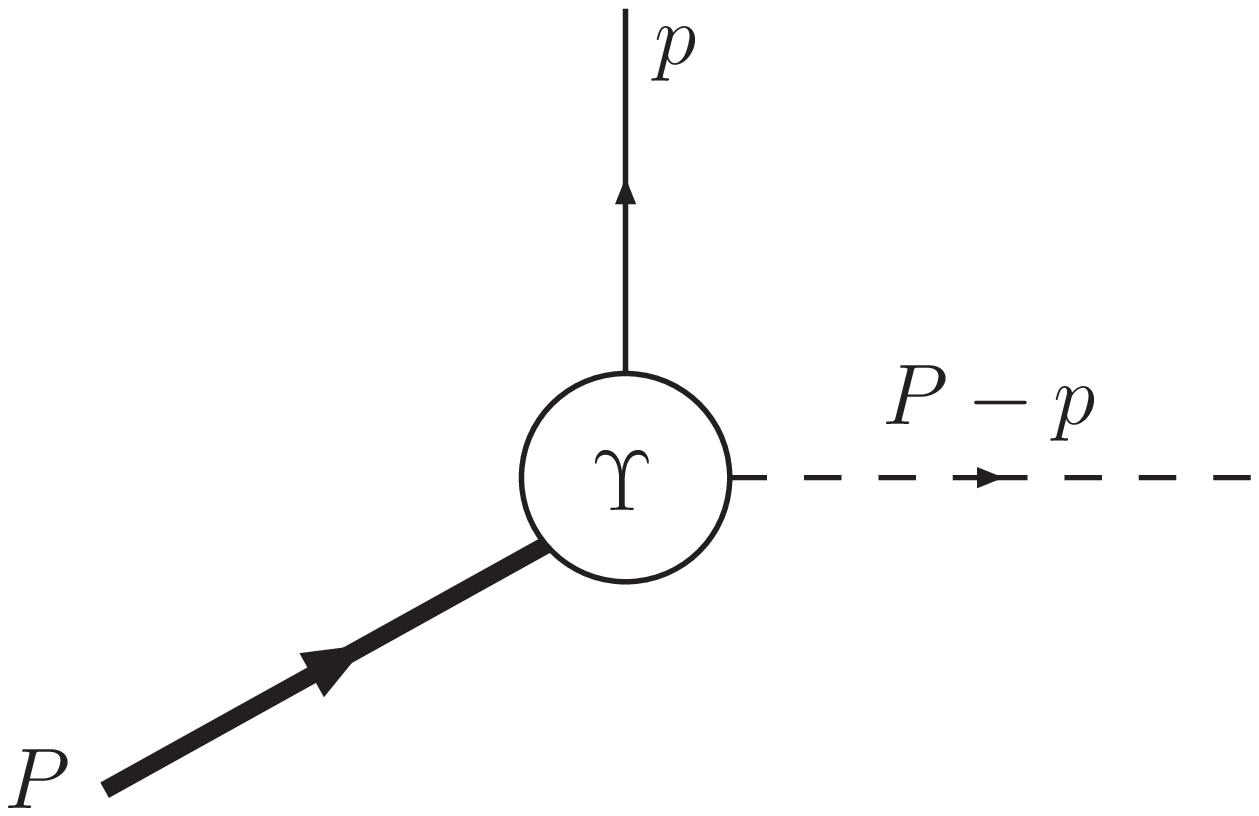}\hspace{2.0cm}
\includegraphics[scale=0.4]{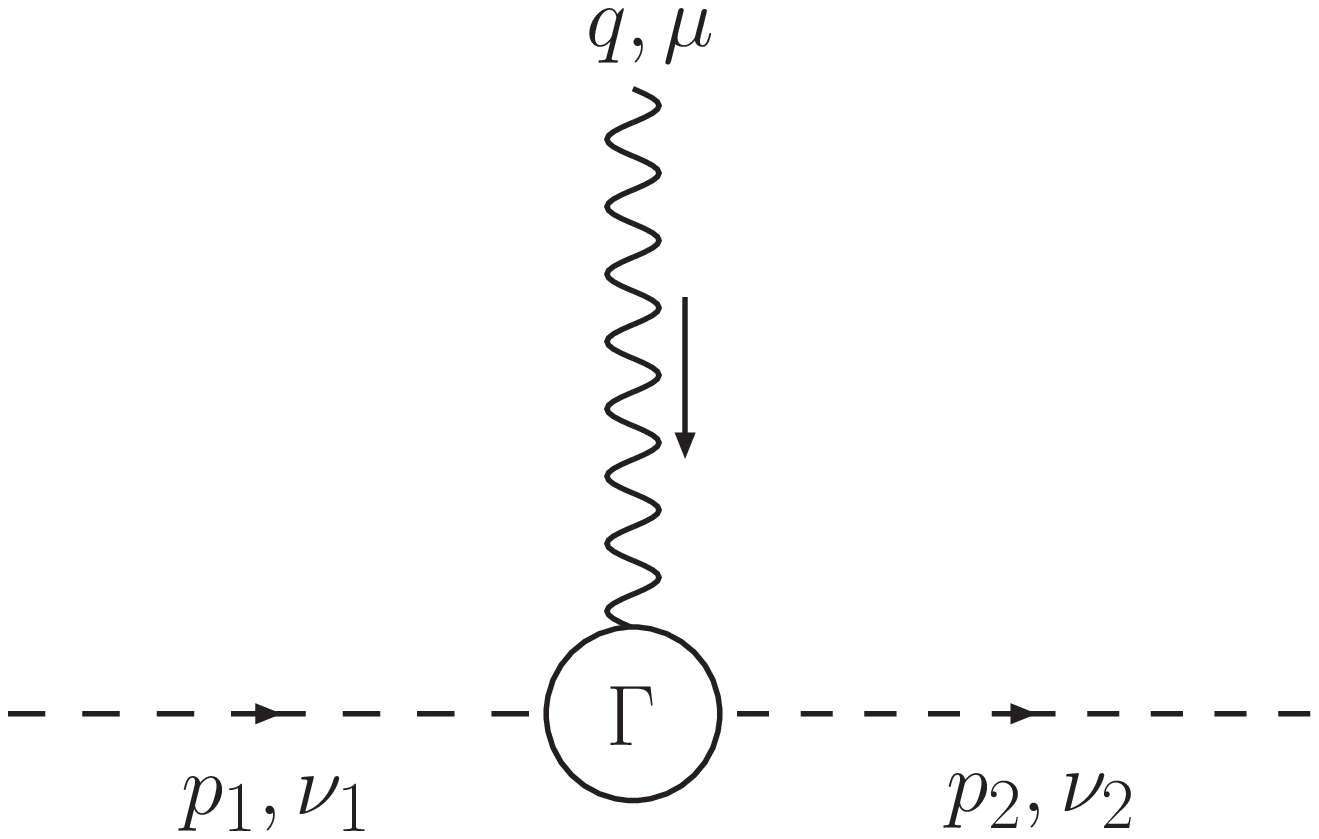}
\caption{\small{Different vertices for the axial-vector diquark.  Left Panel: 
Nucleon-diquark-quark vertex. Right Panel:  Diquark-gluon vertex.}
 \label{cap:Different-vertices-for}}
\end{center}
\end{figure}

For ``T-even'' parton distributions such as the unpolarized PDF $f_{1}$
one obtains a non-vanishing result at leading order (in the nucleon-diquark-quark
coupling) with a trivial contribution from the Wilson line, i.e. at
tree level. In this case the matrix element $\langle dq|\psi|P\rangle$
is depicted in the  Left Panel of 
Fig. \ref{cap:Different-vertices-for}. For a scalar
and an axial-vector diquark different vertices have to be chosen.
The most general nucleon-diquark-quark vertices for off-shell particles
were presented in Ref. \cite{Melnitchouk:1993nk}. For the matrix
elements $\langle dq|\psi|P\rangle$, the nucleon is on-shell which
reduces the amount of structures of the vertices of Ref. \cite{Melnitchouk:1993nk}.
In the following we work with the nucleon-diquark-quark vertices which
were used in Ref. \cite{Jakob:1997wg} to compute ``T-even'' PDFs. They
read for a scalar and an axial-vector diquark\begin{eqnarray}
\Upsilon_{s}(N)=g_{sc}(p^{2}) & ; & \Upsilon_{ax}^{\mu}(N)=\frac{g_{ax}(p^{2})}{\sqrt{3}}\gamma_{5}\left[\gamma^{\mu}-R_{g}\frac{P^{\mu}}{M}\right].\end{eqnarray}
 $g(p^{2})$ are form factors depending on the quark momentum $p$.
They are introduced to yield a more realistic description of the non-perturbative
nature of the quark-quark correlator, and are specified below. $R_{g}$
is a ratio of coupling constants, since both structures in the nucleon-(axial-vector)
diquark-quark coupling can in principle have different couplings.%

To leading order, the matrix elements are given by the following expressions
for a scalar and axial vector diquark, \begin{eqnarray}
\langle sdq;\, P-p|\,\psi_{i}(0)\,|P,S\rangle & = & ig_{sc}(p^{2})\frac{\left[(\slash p+m_{q})u(P,S)\right]_{i}}{p^{2}-m_{q}^{2}+i0},\label{eq:f1MEsc}\\
\langle adq;\, P-p;\lambda|\,\psi_{i}(0)\,|P,S\rangle & = & i\frac{g_{ax}(p^{2})}{\sqrt{3}}\varepsilon_{\mu}^{*}(P-p;\lambda)\frac{\left[(\slash p+m_{q})\gamma_{5}\left[\gamma^{\mu}-R_{g}\frac{P^{\mu}}{M}\right]u(P,S)\right]_{i}}{p^{2}-m_{q}^{2}+i0},\nn\label{eq:f1MEax}\end{eqnarray}
 where the polarization vector of the axial-vector diquark is given
by $\varepsilon_{\mu}$, $u(P,S)$ denotes the nucleon spinor, and
$M$ and $m_{q}$ are nucleon and quark masses, respectively. In this
paper we consider the diquark as a particle with mass $m_{s}$, and
the polarization sum for the axial-vector diquark is\begin{equation}
\sum_{\lambda}\varepsilon_{\mu}^{*}(P-p;\lambda)\varepsilon_{\nu}(P-p;\lambda)=-g_{\mu\nu}+\frac{(P-p)_{\mu}(P-p)_{\nu}}{m_{s}^{2}}.\end{equation}
 The unpolarized TMD $f_{1}$ is obtained by  inserting Eqs.
(\ref{eq:f1MEsc}) and (\ref{eq:f1MEax}) into Eq. (\ref{eq:Correlator})
and projecting $f_{1}$ from the quark-quark correlator (see e.g.
\cite{Goeke:2005hb,Bacchetta:2006tn})\begin{equation}
2f_{1}(x,\vec{p}_{T}^{2})=\frac{1}{2}\int dp^{-}\,\left(\Tr\left[\gamma^{+}\Phi(p;P,S)\right]+\Tr\left[\gamma^{+}\Phi(p;P,-S)\right]\right)\bigg|_{p^{+}=xP^{+}},\end{equation}
 where the {}``$+$'' sign of the $\gamma$-matrix denotes the usual
light cone component ($a^{\pm}=1/\sqrt{2}(a^{0}\pm a^{3})$). The
results for $f_{1}$ in the scalar and axial vector diquark sectors
read \begin{eqnarray}
f_{1}^{sc}(x,\vec{p}_{T}^{2}) & = & \frac{1}{2(2\pi)^{3}}|g_{sc}(p^{2})|^{2}\frac{(1-x)}{\left[\vec{p}_{T}^{2}+\tilde{m}^{2}\right]^{2}}\left[\vec{p}_{T}^{2}+\left(xM+m_{q}\right)^{2}\right],\label{f1sc}\end{eqnarray}
\begin{eqnarray}
f_{1}^{ax}(x,\vec{p}_{T}^{2}) & = & \frac{1}{6(2\pi)^{3}}\frac{|g_{ax}(p^{2})|^{2}}{M^{2}m_{s}^{2}(1-x)\left[\vec{p}_{T}^{2}+\tilde{m}^{2}\right]^{2}}\mathcal{R}_{1}^{ax}\left(x,\vec{p}_{T}^{2};R_{g},\{\mathcal{M}\}\right),\label{f1ax}\end{eqnarray}
 where $\tilde{m}^{2}\equiv xm_{s}^{2}-x(1-x)M^{2}+(1-x)m_{q}^{2}$.
To shorten the notation we introduce a function $\mathcal{R}_1^{ax}$ depending
on $x$ and $\vec{p}_{T}$, and the model parameters $R_{g}$ and
the set of masses, i.e. $\{\mathcal{M}\}\equiv\{ M,\, m_{s},\, m_{q}\}$,
to be fixed below. 


Another ``T-even'' function of interest is the distribution of transversely
polarized quarks in a longitudinally polarized target, 
\begin{eqnarray}
2\lambda_{P}p_{T}^{i}h_{1L}^{\perp}(x,\vec{p}_{T}^{2}) = \frac{M}{2}\int dp^{-}\,\left\{ \Tr\left[\gamma^{+}\gamma^{i}\gamma_{5}\Phi(p;P,S_{L})\right]-\Tr\left[\gamma^{+}\gamma^{i}\gamma_{5}\Phi(p;P,-S_{L})\right]\right\} ,
\end{eqnarray}
 where $\lambda_{P}$ is the target helicity, and $S_{L}$ is the
spin 4-vector in longitudinal direction, i.e. $S_{L}=\left[-\frac{\lambda_{P}}{M}P^{-},\frac{\lambda_{P}}{M}P^{+},\vec{0}_{T}\right]$.
By applying the same methods as for $f_{1}$, we obtain\begin{eqnarray}
h_{1L}^{\perp,sc}(x,\vec{p}_{T}^{2})  =  -\frac{|g_{sc}(p^{2})|^{2}}{(2\pi)^{3}}\frac{(1-x)M(xM+m_{q})}{\left[\vec{p}_{T}^{2}+\tilde{m}^{2}\right]^{2}},\label{h1Lperpsc}\end{eqnarray}
\begin{eqnarray}
h_{1L}^{\perp,ax}(x,\vec{p}_{T}^{2})  =  \frac{|g_{ax}(p^{2})|^{2}}{12(2\pi)^{3}}\frac{1}{\left[\vec{p}_{T}^{2}+\tilde{m}^{2}\right]^{2}Mm_{s}^{2}(1-x)}\mathcal{R}_{1L}^{\perp,ax}\left(x,\vec{p}_{T}^{2};R_{g},\{\mathcal{M}\}\right),\label{h1Lperpax}\end{eqnarray}
where  for brevity  $\mathcal{R}_1^{ax}$ and 
$\mathcal{R}_{1L}^{\perp,ax}$ are given 
in Appendix \ref{append3}.

\begin{figure}
\begin{center}\includegraphics[scale=0.4]{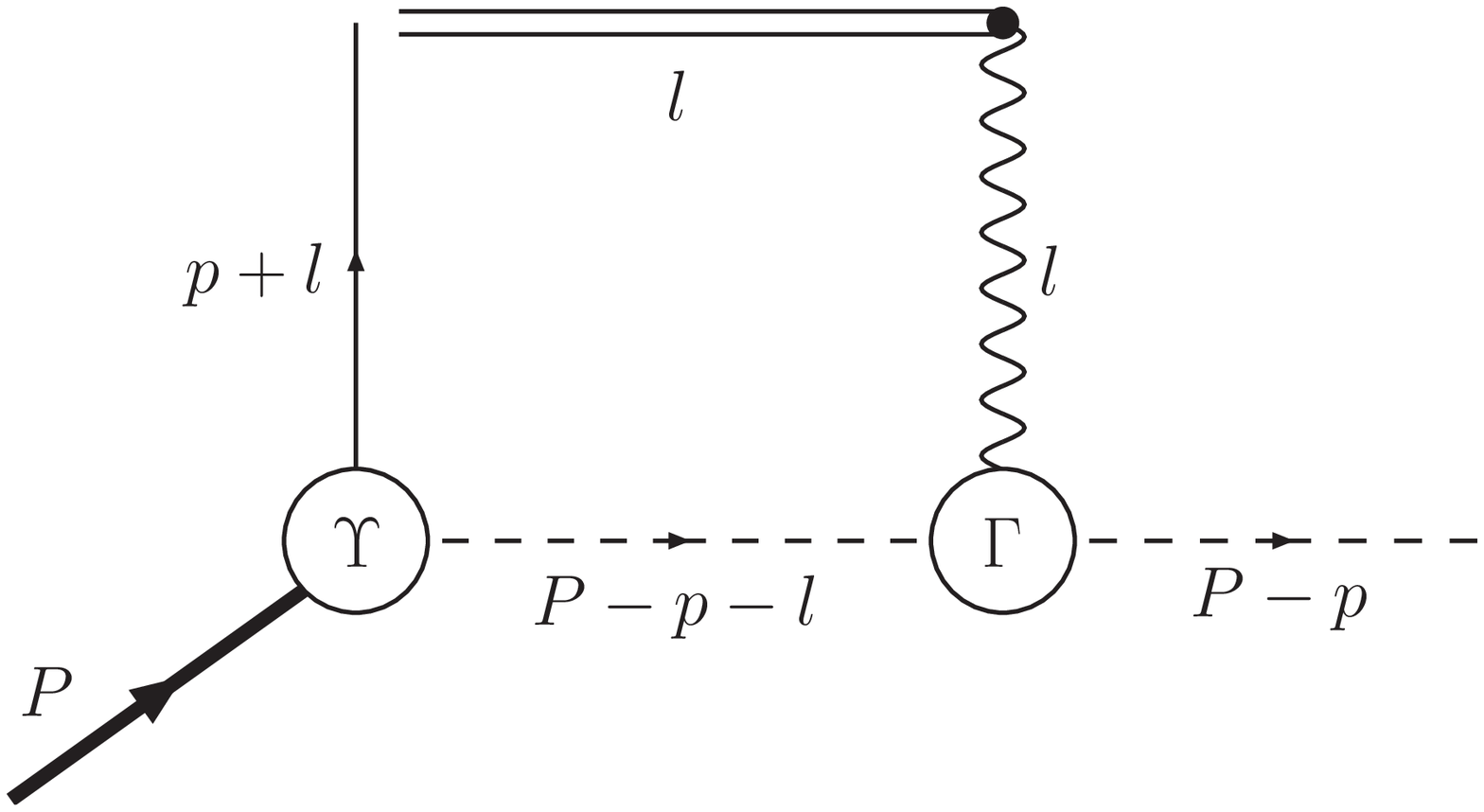}\hspace{0.25cm}
~~\includegraphics[scale=0.4]{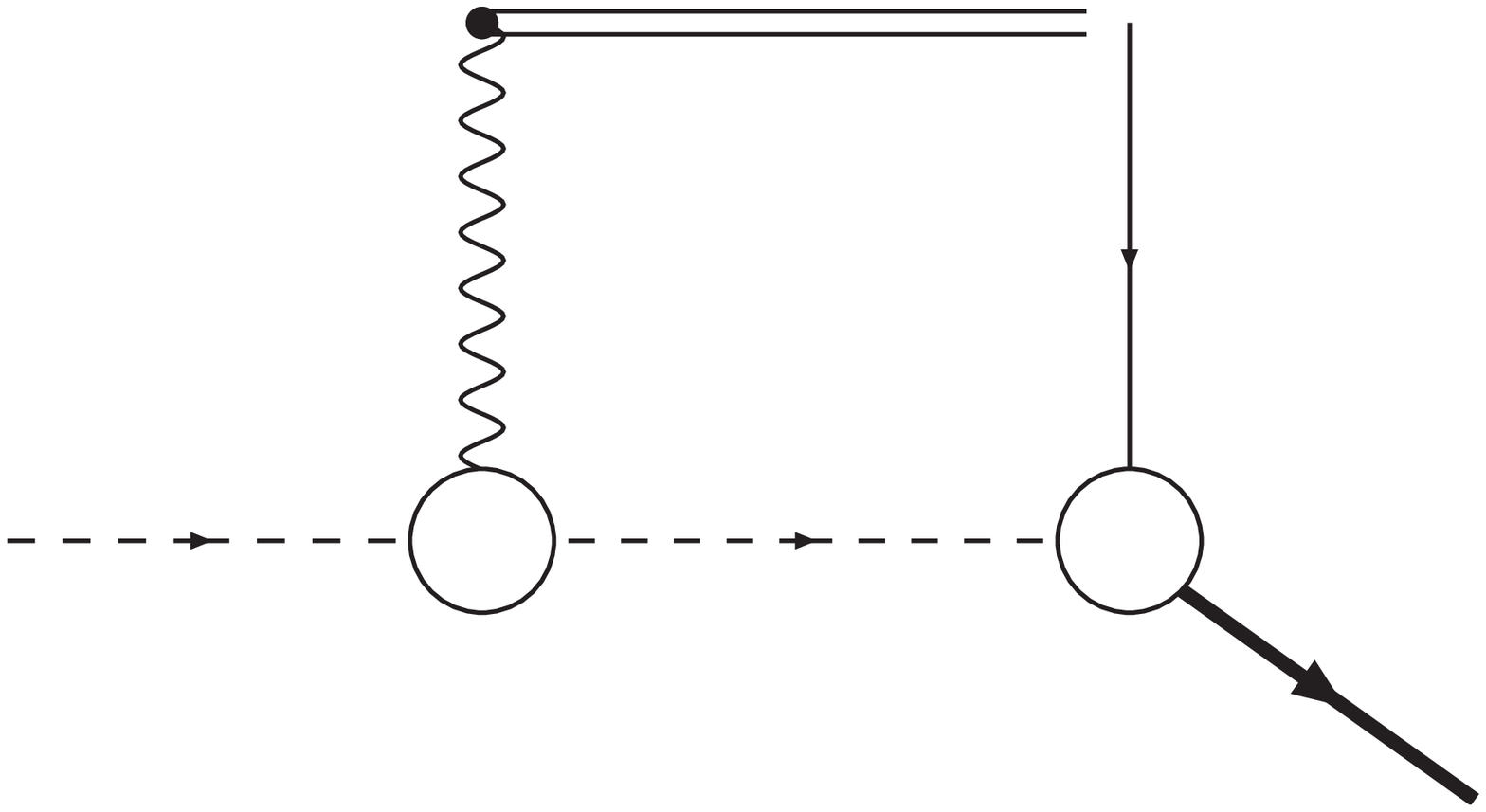}
\caption{\small{Contribution of the gauge link in the one-gluon approximation.  Left Panel: Box-graph. Right Panel: Box-graph hermitian conjugated.}\label{cap:Contribution-of-the}}
\end{center}
\end{figure}

By contrast, ``T-odd'' PDFs cannot be generated by simply considering
the tree-level diagram in the 
Left Panel  of  Fig. \ref{cap:Different-vertices-for}.
In the spectator framework the ``T-odd'' PDFs \cite{Brodsky:2002cx} 
are generated by the gauge link  in 
Eq.~(\ref{eq:Fully-unQQ})~\cite{Ji:2002aa,
Goldstein:2002vv,Gamberg:2003ey,Boer:2002ju}.
Thus, the leading contribution can be obtained by expanding the exponential
of the gauge link up to first order. This contribution results in
a box diagram as 
shown in the Left Panel of Fig. \ref{cap:Contribution-of-the}, 
which contains an imaginary part necessary for ``T-odds''. We restrict
ourselves to the case where one gluon models the final state interactions.
The contribution of the gauge link is represented in the Left Panel of
Fig. \ref{cap:Contribution-of-the}
 by the double (eikonal) line and the eikonal vertex yielding a
contribution to the box diagram \begin{equation}
\frac{i}{\left[l\cdot v+i0\right]}\times(-ie_{q}v^{\lambda}),\end{equation}
 where $l$ is the loop momentum, $e_{q}$ the charge of the quark
and $v$ is a light cone vector representing the direction of the
Wilson line. In order to evaluate the box diagram we need to specify
the gluon-diquark coupling. With an one-gluon exchange approximation 
in mind we use the gluon-diquark coupling for
a scalar diquark, and for an axial-vector diquark
we use a general axial-vector-vector that  models the composite 
nature of the diquark through an anomalous magnetic
moment $\kappa$~\cite{Goldstein:1979wb}. 
In the notations of Fig. \ref{cap:Different-vertices-for}
(Right Panel) the gluon-diquark vertices read\begin{equation}
\Gamma_{s}^{\mu}=-ie_{dq}(p_{1}+p_{2})^{\mu},\end{equation}
\begin{equation}
\Gamma_{ax}^{\mu\nu_{1}\nu_{2}}=-ie_{dq}\left[g^{\nu_{1}\nu_{2}}(p_{1}+p_{2})^{\mu}+(1+\kappa)\left(g^{\mu\nu_{2}}(p_{2}+q)^{\nu_{1}}+g^{\mu\nu_{1}}(p_{1}-q)^{\nu_{2}}\right)\right].\end{equation}
 For $\kappa=-2$ the vertex $\Gamma_{ax}$ reduces to the standard
$\gamma WW$-vertex. We can now express the matrix elements
including the gauge link in the one gluon approximation in the following
way 
\begin{eqnarray}
 && \hspace{-1cm}\left.\langle sdq;\, P-p|\,\mathcal{W}[\infty,0,\vec{0}_{T}\,|\,0]\,\psi_{i}(0)\,|P,S\rangle\right|_{1-gl}
\nn
&=& \hspace{-0.25cm} -ie_{q}e_{dq}\int\frac{d^{4}l}{(2\pi)^{4}}\; g_{sc}((l+p)^{2})\mathcal{D}^{sc}(P-p-l)\frac{\left[(\slash p+\slash l+m_{q})u(P,S)\right]_{i}v\cdot(2P-2p-l)}{\left[l\cdot v+i0\right]\left[l^{2}+i0\right]\left[(l+p)^{2}-m_{q}^{2}+i0\right]},\nn
\label{eq:BMMEsc}\end{eqnarray}
\begin{eqnarray}
& & \hspace{-2cm} \left.\langle adq;\, P-p,\lambda|\,\mathcal{W}[\infty,0,\vec{0}_{T}\,|\,0]\,\psi_{i}(0)\,|P,S\rangle\right|_{1-gl}\nonumber \\
& =& -ie_{q}e_{dq}\int\frac{d^{4}l}{(2\pi)^{4}}\;\frac{g_{ax}\left((p+l)^{2}\right)}{\sqrt{3}}\varepsilon_{\sigma}^{*}(P-p,\lambda)\mathcal{D}_{\rho\eta}^{ax}(P-p-l)\times\nonumber \\
 && \frac{\left[g^{\sigma\rho}\, v\cdot(2P-2p-l)+(1+\kappa)\left(v^{\sigma}\,(P-p+l)^{\rho}+v^{\rho}\,(P-p-2l)^{\sigma}\right)\right]}{\left[l\cdot v+i0\right]\left[l^{2}+i0\right]\left[(l+p)^{2}-m_{q}^{2}+i0\right]}\times\nonumber \\
 & &\left[\left(\slash p+\slash l+m_{q}\right)\gamma_{5}\left(\gamma^{\eta}-R_{g}\frac{P^{\eta}}{M}\right)u(P,S)\right]_{i}, \label{eq:BMMEax}\end{eqnarray}
where the subscript, $1-gl$ denotes ``one gluon exchange''. 
 In these expressions $\mathcal{D}$ denotes the propagator of the
scalar and axial-vector diquark,\begin{equation}
\mathcal{D}^{sc}(P-p-l)=\frac{i}{\left[(P-p-l)^{2}-m_{s}^{2}+i0\right]},\end{equation}
\begin{equation}
\mathcal{D}_{\mu\nu}^{ax}(P-p-l)=\frac{-i\left(g_{\mu\nu}-\frac{(P-p-l)_{\mu}(P-p-l)_{\nu}}{m_{s}^{2}}\right)}{\left[(P-p-l)^{2}-m_{s}^{2}+i0\right]}.\label{eq:axProp}\end{equation}
 The term $\frac{(P-p-l)_{\mu}(P-p-l)_{\nu}}{m_{s}^{2}}$ is a crucial
difference of our approach compared to the calculation in Ref. \cite{Bacchetta:2003rz},
where the dependence on the proton and spectator momenta 
inside loop integral is absent. 
It is shown below that this leads to various complications when performing
the loop-integral. %

In the similar fashion as for $f_{1}$ and $h_{1L}^{\perp}$, we extract
the Boer-Mulders function by inserting Eqs. (\ref{eq:BMMEsc}) and
(\ref{eq:BMMEax}) (and the tree-matrix elements (\ref{eq:f1MEsc})
and (\ref{eq:f1MEax}), i.e. the leading non-trivial perturbative
contribution is the interference term between tree-graph and box-graph)
into the quark-quark correlator (\ref{eq:Correlator})
\begin{equation}
2\epsilon_{T}^{ij}p_{T}^{j}h_{1}^{\perp}(x,\vec{p}_{T}^{2})=\frac{M}{2}\int dp^{-}\left(\Tr\left[\Phi_{\mathrm{unpol}}(p,S)i\sigma^{i+}\gamma_{5}\right]+\Tr\left[\Phi_{\mathrm{unpol}}(p,-S)i\sigma^{i+}\gamma_{5}\right]\right)\bigg|_{p^{+}=xP^{+}}, \label{bm}\end{equation}
where $\epsilon_{T}^{ij}\equiv\epsilon^{-+ij}$ and $\epsilon^{0123}=+1$.

\section{Boer-Mulders function for an axial-vector diquark}

We proceed with calculating the Boer-Mulders function in the axial-vector
diquark sector. As described above, the interference term between
tree- and box graph reads\begin{eqnarray}
\epsilon_{T}^{ij}p_{T}^{j}h_{1}^{\perp,ax}(x,\vec{p}_{T}^{2}) & = & -\frac{e_{q}e_{dq}}{8(2\pi)^{3}}\frac{1}{\vec{p}_{T}^{2}+\tilde{m}^{2}}\frac{M}{P^{+}}\;\int\frac{d^{4}l}{(2\pi)^{4}}\;\bigg\{\frac{1}{3}g_{ax}\left((l+p)^{2}\right)g_{ax}^{*}\left(p^{2}\right)\times\nonumber \\
 &  & \hspace{-2cm}\mathcal{D}_{\rho\eta}(P-p-l)\left(\sum_{\lambda}\varepsilon_{\sigma}^{*}(P-p;\lambda)\varepsilon_{\mu}(P-p;\lambda)\right)\times\nonumber \\
 &  & \hspace{-2cm}\frac{\left[g^{\sigma\rho}\, v\cdot(2P-2p-l)+(1+\kappa)\left(v^{\sigma}\,(P-p+l)^{\rho}+v^{\rho}\,(P-p-2l)^{\sigma}\right)\right]}{\left[l\cdot v+i0\right]\left[l^{2}-\lambda^{2}+i0\right]\left[(l+p)^{2}-m_{q}^{2}+i0\right]}\times\nonumber \\
 &  & \hspace{-3cm} \Tr\Big[\left(\slash P+M\right)\left(\gamma^{\mu}-R_{g}\frac{P^{\mu}}{M}\right)\left(\slash p-m_{q}\right)\gamma^{+}\gamma^{i}\times
\left(\slash l+\slash p+m_{q}\right)\left(\gamma^{\eta}+R_{g}\frac{P^{\eta}}{M}\right)\gamma_{5}\Big]\bigg\}+\;\mathrm{h.c.}\ .\label{eq:StartingFormula}
\nn
\end{eqnarray}
 The momentum of the quark, $p$ is specified by \begin{equation}
p=\left[p^{-}=-\frac{\vec{p}_{T}^{2}+m_{s}^{2}-(1-x)M^{2}}{2(1-x)P^{+}},\, p^{+}=xP^{+},\,\vec{p}_{T}\right].\end{equation}

\subsection{Light cone integration}

Here we comment on 
the evaluation of the 
four-dimensional loop integral in Eq.~\ref{eq:StartingFormula}. 
A convenient way to simplify the calculation is to sort the numerator
in terms of loop momenta and consider each term separately. Since
the numerator in Eq. (\ref{eq:StartingFormula}) contains at most
the loop momentum to the power of four we can write it in the following
way\begin{equation}
\mathrm{numerator}=\sum_{i=1}^{4}N_{\alpha_{1}...\alpha_{i}}^{(i)}l^{\alpha_{1}}...l^{\alpha_{i}}+N^{(0)}.\label{eq:NumDecomp}\end{equation}
 The (real) coefficients (tensors) $N_{\alpha_{1}...\alpha_{i}}^{(i)}$
depend only on external momenta $P$ (nucleon momentum) and $p$ (quark
momentum) and can be computed in a straight-forward but tedious calculation.
We used the \emph{Mathematica}-package TRACER \cite{Jamin:1991dp}
for this decomposition (\ref{eq:NumDecomp}). The advantage of this
procedure is that we are left with an arbitrary integral of the form\begin{equation}
J^{(i)\alpha_{1}\alpha_{2}...\alpha_{i}}\equiv\int\frac{d^{4}l}{(2\pi)^{4}}\;\frac{\frac{1}{3}g_{ax}((l+p)^{2})g_{ax}^{*}(p^{2})l^{\alpha_{1}}l^{\alpha_{2}}...l^{\alpha_{i}}}{\left[l\cdot v+i0\right]\left[l^{2}-\lambda^{2}+i0\right]\left[(l+p-P)^{2}-m_{s}^{2}+i0\right]\left[(l+p)^{2}-m_{q}^{2}+i0\right]},\label{eq:MasterIntegral}\end{equation}
 and the light cone components of the loop momentum, $l^{+}$ and
$l^{-}$, can be integrated out easily.

We sketch the light cone integration. First, we specify the vector
$v$ to be a light cone vector $v=[v^{-}=1,v^{+}=0,\vec{v}_{T}=0]$
representing the Wilson line. Thus, the product $l\cdot v$ reduces
to $l^{+}$ and doesn't contribute to the $l^{-}$ integration. Next,
we perform the integral over $l^{-}$ via contour integration and
encounter three poles in the $l^{-}$-plane from the last three terms
in the denominator in Eq. (\ref{eq:MasterIntegral}). The integral
is non-vanishing when $-xP^{+}<l^{+}<(1-x)P^{+}$, otherwise all poles
are located in the same complex $l^{-}$ half-plane. For $-xP^{+}<l^{+}<(1-x)P^{+}$
the third factor in the denominator in Eq. (\ref{eq:MasterIntegral})
has always a positive imaginary part while the forth factor has always
a negative one. The imaginary part of the second factor becomes positive
for $l^{+}<0$ and negative for $l^{+}>0$. We close the contour of
integration in the upper half-plane which excludes the forth factor
in the denominator, and the second for $l^{+}>0$. 
Thus, we obtain
{\small
\begin{eqnarray}
 & & J^{(i)\alpha_{1}...\alpha_{i}}=\frac{1}{3}i\int\frac{d^{2}\vec{l}_{T}}{(2\pi)^{2}}\int_{-xP^{+}}^{(1-x)P^{+}}\frac{dl^{+}}{2\pi}\;\frac{1}{\left[l^{+}+i0\right]}\times\\
& & \frac{1}{\left[2l^{+}\left(2(l^{+}-(1-x)P^{+})(P^{-}-p^{-})+[(\vec{l}_{T}+\vec{p}_{T})^{2}+m_{s}^{2}]\right)-2(l^{+}-(1-x)P^{+})[\vec{l}_{T}^{2}+\lambda^{2}]\right]}\times\nonumber \\
 & &\bigg\{\frac{\left[2(l^{+}-(1-x)P^{+})\right]\left[g_{ax}((l+p)^{2})g_{ax}^{*}(p^{2})(l^{\alpha_{1}}...l^{\alpha_{i}})\right]\bigg|_{l^{-}=P^{-}-p^{-}+\frac{(\vec{l}_{T}+\vec{p}_{T})^{2}+m_{s}^{2}}{2(l^{+}-(1-x)P^{+})}}}{\left[2(l^{+}-(1-x)P^{+})\left(2(l^{+}+xP^{+})P^{-}-[(\vec{l}_{T}+\vec{p}_{T})^{2}+m_{q}^{2}]\right)+2(l^{+}+xP^{+})[(\vec{l}_{T}+\vec{p}_{T})^{2}+m_{s}^{2}]\right]}\nonumber \\
 & & -\frac{\left[2l^{+}\right]\Theta(-l^{+})\left[g_{ax}((l+p)^{2})g_{ax}^{*}(p^{2})(l^{\alpha_{1}}...l^{\alpha_{i}})\right]\bigg|_{l^{-}=\frac{\vec{l}_{T}^{2}+\lambda^{2}}{2l^{+}}}}{\left[2l^{+}\left(2(l^{+}+xP^{+})p^{-}-[(\vec{l}_{T}+\vec{p}_{T})^{2}+m_{q}^{2}]\right)+2(l^{+}+xP^{+})[\vec{l}_{T}^{2}+\lambda^{2}]\right]}\bigg\}.\nonumber \end{eqnarray}}
We calculate the $l^{+}$-integral by adding the
complex conjugated integral (stemming from the complex conjugated
interference graph). Since $l^{+}+i0$ is the only propagator remaining
with an imaginary part, adding the complex conjugated integral results
into a $\delta$-function contribution via the relation\begin{equation}
\frac{1}{l^{+}+i0}-\frac{1}{l^{+}-i0}=-2\pi i\delta(l^{+}).\end{equation}
 We obtain\begin{eqnarray}
& & \hspace{-1cm}J^{(i)\alpha_{1}...\alpha_{i}}+\left(J^{(i)\alpha_{1}...\alpha_{i}}\right)^{*}\label{eq:LCint}\\
 & =&\frac{1}{3}\int\frac{d^{2}\vec{l}_{T}}{(2\pi)^{2}}\;\bigg\{\frac{(-1)\left(g_{ax}((l+p)^{2})g_{ax}^{*}(p^{2})\left[l^{\alpha_{1}}...l^{\alpha_{i}}\right]\right)\bigg|_{l^{-}=\frac{\vec{p}_{T}^{2}-(\vec{l}_{T}+\vec{p}_{T})^{2}}{2(1-x)P^{+}},\; l^{+}=0}}{2P^{+}\left[\vec{l}_{T}^{2}+\lambda^{2}\right]\left[(\vec{l}_{T}+\vec{p}_{T})^{2}+\tilde{m}^{2}\right]}\nonumber \\
 && -\int_{-xP^{+}}^{(1-x)P^{+}}dl^{+}\;\Big[\frac{\left[\frac{l^{+}}{xP^{+}}\right]\Theta(-l^{+})\delta(l^{+})\left(g_{ax}((l+p)^{2})g_{ax}^{*}(p^{2})\left[l^{\alpha_{1}}...l^{\alpha_{i}}\right]\right)\bigg|_{l^{-}=\frac{\vec{l}_{T}^{2}+\lambda^{2}}{2l^{+}}}}{\left[2(1-x)P^{+}[\vec{l}_{T}^{2}+\lambda^{2}]\right]\left[\vec{l}_{T}^{2}+\lambda^{2}\right]}\Big]\bigg\}.\nonumber \end{eqnarray}

At this point we are forced to specify the form factor $g_{ax}$ since
the second integral in Eq. (\ref{eq:LCint}) is potentially ill-defined.
This happens when $g(p^{2})$ is a holomorphic function in $p^{2}$
(i.e. it contains no poles) and at least one of the Minkowski indices
is light-like in the minus direction, e.g. $\alpha_{1}=-$, $\alpha_{2},...,\alpha_{i}\in\left\{ +,\perp\right\} $.
In such cases we end up with an integral of the form $\int dl^{+}\,\delta(l^{+})\Theta(-l^{+})$,
which is ill-defined. This implies that $l^{+}=0$ and $l^{-}=\infty$
which signals the existence of a light cone divergence as was shown
in Ref. \cite{Gamberg:2006ru}. While for a \emph{scalar} diquark
one doesn't encounter a Minkowski-index, $\alpha_{j}=-$, for twist-2
``T-odd'' PDFs such as the Boer-Mulders function (so that there are no
light cone divergences in this case), it was shown in Ref. \cite{Gamberg:2006ru}
that for twist-3 ``T-odd'' PDFs light cone divergences exist for a scalar
diquark. However, calculating the 
coefficients $N_{\alpha_{1}...\alpha_{i}}^{(i)}$
in Eq. (\ref{eq:NumDecomp}) for an axial-vector diquark, one of the
Minkowski-indices can be a {}``minus'' (i.e. $\alpha_{j}=-$). Thus,
for an axial-vector diquark we encounter a light cone divergence already
for twist-2 ``T-odd'' PDFs. Here it is worth mentioning that such divergences
do not arise in a pQCD-quark-target model where the spectator state
is a gluon \cite{Goeke:2006ef}.

From the standpoint of phenomenology one can regard these light cone
divergences as model artifacts (for the axial-vector diquark model).
It was shown in Ref. \cite{Gamberg:2006ru} how to regularize these
light cone divergences by introducing non-light-like Wilson line.
It was pointed out that one can also handle the light cone divergences
by introducing phenomenological form factors with additional poles.
Like the quark propagator in Eq. (\ref{eq:StartingFormula}) they
introduce additional factors of $l^{+}$ in the numerator of the second
term in Eq. (\ref{eq:LCint}). We adopt this procedure to model the
Boer-Mulders function and choose a form factor of the following form\begin{equation}
g_{ax}(p^{2})=N_{ax}^{n-1}\frac{(p^{2}-m_{q}^{2})f(p^{2})}{\left[p^{2}-\Lambda^{2}+i0\right]^{n}},\label{eq:FFn}\end{equation}
 and find that for $n\geq3$ (for $n\geq2$ is already sufficient
for $\kappa=-2$ ) enough powers of $l^{+}$ enter the numerator of
the second term in Eq. (\ref{eq:LCint}) to compensate the minus components
$l^{-}$. The second term then vanishes since $\int dl^{+}\, l^{+n}\delta(l^{+})\Theta(-l^{+})=0$
for $n\geq1$. In Eq. (\ref{eq:FFn}) $f(p^{2})$ is then a function
without poles to be fixed below, while $\Lambda$ is an arbitrary
mass scale to be fixed by phenomenology (i.e. fitting $f_{1}$ to
data). $N_{ax}$ is a normalization factor.

After performing the light cone integrations we are then left with
the remaining integral over the transverse loop momentum (for $n=3$)
\begin{eqnarray}
\epsilon_{T}^{ij}p_{T}^{j}h_{1}^{\perp,ax}(x,\vec{p}_{T}^{2}) & = & -\frac{e_{q}e_{dq}}{8(2\pi)^{3}}N_{ax}^{4}\frac{1}{3}\frac{(1-x)^{3}f(p^{2})}{\left[\vec{p}_{T}^{2}+\tilde{m}_{\Lambda}^{2}\right]^{3}m_{s}^{4}}\times\nonumber \\
 &  & \int\frac{d^{2}\vec{l}_{T}}{(2\pi)^{2}}\;\frac{f^{*}((p+l)^{2})}{\left[\vec{l}_{T}^{2}+\lambda^{2}\right]\left[(\vec{l}_{T}+\vec{p}_{T})^{2}+\tilde{m}_{\Lambda}^{2}\right]^{3}}\times\nonumber \\
 &  & \bigg\{\epsilon_{T}^{ij}p_{T}^{j}\left[(\vec{l}_{T}^{2})^{2}A_{p}+2(\vec{l}_{T}\cdot\vec{p}_{T})\vec{l}_{T}^{2}B_{p}+\vec{l}_{T}^{2}C_{p}+2(\vec{l}_{T}\cdot\vec{p}_{T})D_{p}+E_{p}\right]+\nonumber \\
 &  & \epsilon_{T}^{ij}l_{T}^{j}\left[(\vec{l}_{T}^{2})^{2}A_{l}+2(\vec{l}_{T}\cdot\vec{p}_{T})\vec{l}_{T}^{2}B_{l}+\vec{l}_{T}^{2}C_{l}+2(\vec{l}_{T}\cdot\vec{p}_{T})D_{l}+E_{l}\right]\nonumber \\
 &  & +\epsilon_{T}^{rs}l_{T}^{r}p_{T}^{s}\left[(A_{lp}\vec{l}_{T}^{2}+2\vec{l}_{T}\cdot\vec{p}_{T}B_{lp})(l_{T}^{i}+p_{T}^{i})+E_{lp}(l_{T}^{i}+2p_{T}^{i})\right]\bigg\},\label{eq:Transverse}\end{eqnarray}
 where the coefficients which are functions of $x,M,m_{q},p_{T}^{2},\kappa\,\, and\,\, R_{g}$
are given in the Appendix \ref{append1}. $\tilde{m}_{\Lambda}^{2}$
replaces $\tilde{m}^{2}$ by $\tilde{m}_{\Lambda}^{2}=xm_{s}^{2}-x(1-x)M^{2}+(1-x)\Lambda^{2}$.  
We point out that the vanishing of the coefficient $E_{p}$ \emph{ensures
that no IR-divergence appears in the transverse integral (\ref{eq:Transverse})}.
This serves as an important check of the calculation.

\subsection{Transverse integral}

The transverse integral (\ref{eq:Transverse}) can be calculated in
a straightforward manner. We note that the transverse integral (\ref{eq:Transverse})
is UV-divergent if we chose $f(p^{2})=1$. This can be seen from naive
power counting of the integrand. The UV-divergence stems from the
term $\vec{l}_{T}^{4}$, which in turn is a consequence of the fact
that we took the full numerator of the axial-vector diquark propagator
into account, in contrast to Ref. \cite{Bacchetta:2003rz} where no
UV-divergences were reported. We regularize it by choosing $f(p^{2})$
to be a covariant Gaussian,\begin{equation}
f((l+p)^{2})=\e^{-b|l+p|^{2}},\label{eq:FFn2}\end{equation}
 where $b$ is interpreted as the the regulator of the high $l$ integration.
Due to the pole contribution $l^{+}=0$, the Gaussian has no effect
on the light cone integration. Thus we can write the squared products
of the momenta $l+p$ and $p$ \--- after performing the light cone
integration \--- as follows, \begin{eqnarray}
(l+p)^{2} & = & -\frac{(\vec{l}_{T}+\vec{p}_{T})^{2}+xm_{s}^{2}-x(1-x)M^{2}}{1-x},\\
p^{2} & = & -\frac{\vec{p}_{T}^{2}+xm_{s}^{2}-x(1-x)M^{2}}{1-x}.\end{eqnarray}
 Now all that remains is to perform the $\vec{l}_{T}$ integration.
After a shift of the integration variable from $\vec{l}_{T}\rightarrow\vec{l}_{T}+\vec{p}_{T}$
it is convenient to use polar coordinates to calculate the integral,
and perform the angular integration first. For this we choose a coordinate
system in such a way that the $x$-axis is along $\vec{p}_{T}$, such
that $\vec{p}_{T}=|\vec{p}_{T}|(1,0)$. The integration is performed
with respect to that direction, i.e. $\vec{l}_{T}=|\vec{l}_{T}|(\cos\phi,\sin\phi$).
Having fixed the coordinate system in such a way, the hanging index
$i$ can only be $2$. We perform now the angular integration over
$\phi$ by means of the formula\begin{equation}
\int_{0}^{\pi}\frac{\cos(nx)dx}{1+a\cos(x)}=\frac{\pi}{\sqrt{1-a^{2}}}\left(\frac{\sqrt{1-a^{2}}-1}{a}\right)^{n},\,\,\,\,\,\,\, a^{2}<1,\,\,\,\, n\geq0.\end{equation}
 We are left with the remaining one-dimensional integrals ($\sqrt{z}\equiv l$)
\begin{eqnarray}
h_{1}^{\perp,ax}(x,\vec{p}_{T}^{2}) & = & -\frac{e_{q}e_{dq}}{48(2\pi)^{4}}N_{ax}^{4}\frac{(1-x)^{3}\e^{-\tilde{b}(\vec{p}_{T}^{2}+2xm_{s}^{2}-2x(1-x)M^{2})}}{m_{s}^{4}\left[\vec{p}_{T}^{2}+\tilde{m}_{\Lambda}^{2}\right]^{3}}\times\nonumber \\
 &  & \Bigg(\int_{0}^{\infty}\frac{\e^{-\tilde{b}z}\; dz}{\left[z+\tilde{m}_{\Lambda}^{2}\right]^{3}}\left[z(A_{p}-2A_{l}+B_{l})+\vec{p}_{T}^{2}(A_{p}-A_{l}-2(B_{p}-B_{l}))+C_{p}-C_{l}\right]\nonumber \\
 &  & +\int_{0}^{\vec{p}_{T}^{2}}\frac{\e^{-\tilde{b}z}\; dz}{\left[z+\tilde{m}_{\Lambda}^{2}\right]^{3}}\left[z^{2}\frac{A_{lp}}{2\vec{p}_{T}^{2}}+z\left(-\frac{1}{2}A_{lp}-\frac{D_{l}}{\vec{p}_{T}^{2}}+\frac{E_{lp}}{2\vec{p}_{T}^{2}}\right)-2(D_{p}-D_{l})-\frac{E_{l}}{\vec{p}_{T}^{2}}\right]\nonumber \\
 &  & +\int_{\vec{p}_{T}^{2}}^{\infty}\frac{\e^{-\tilde{b}z}\; dz}{\left[z+\tilde{m}_{\Lambda}^{2}\right]^{3}}\left[z\frac{A_{lp}}{2}-\vec{p}_{T}^{2}\frac{A_{lp}}{2}+D_{l}+\frac{1}{2}E_{lp}\right]\Bigg),\end{eqnarray}
 where $\tilde{b}\equiv b/(1-x)$. These integrals can be expressed
in terms of incomplete $\Gamma$-functions\begin{equation}
\Gamma(n,x)\equiv\int_{x}^{\infty}\e^{-t}t^{n-1}dt.\end{equation}
The Boer-Mulders function for an axial-vector diquark then reads\begin{eqnarray}
h_{1}^{\perp,ax}(x,\vec{p}_{T}^{2}) & = & -\frac{e_{q}e_{dq}}{48(2\pi)^{4}}N_{ax}^{4}\frac{(1-x)^{3}\e^{-\tilde{b}\vec{p}_{T}^{2}-2\tilde{b}(xm_{s}^{2}-x(1-x)M^{2})}}{m_{s}^{4}\left[\vec{p}_{T}^{2}+\tilde{m}_{\Lambda}^{2}\right]^{3}}\mathcal{R}_{1}^{\perp,ax}\left(x,\vec{p}_{T}^{2};R_{g},\kappa,\tilde{b},\Lambda,\{\mathcal{M}\}\right),\nonumber\\
& &  \label{eq:BMax} \end{eqnarray}
where the explicit form of $\mathcal{R}_{1}^{\perp}$ is expressed
in term of incomplete Gamma functions and can be found in Appendix
\ref{append3}.

The Boer Mulders-function for a scalar diquark is much easier to calculate
\cite{Goldstein:2002vv} due to its simpler Dirac-trace structure.
The light-cone divergences we have encountered in the axial-vector
diquark sector do not appear for the scalar sector. With our choice
of the form factor (cf. Eqs.~(\ref{eq:FFn}) and (\ref{eq:FFn2})) the
Boer-Mulders function for a scalar diquark reads \begin{eqnarray}
h_{1}^{\perp,sc}(x,\vec{p}_{T}^{2}) & = & \frac{e_{q}e_{dq}}{4(2\pi)^{4}}N_{sc}^{4}\frac{(1-x)^{5}M(xM+m_{q})}{\vec{p}_{T}^{2}\left[\vec{p}_{T}^{2}+\tilde{m}_{\Lambda}^{2}\right]^{3}}\e^{-\tilde{b}(\vec{p}_{T}^{2}+2xm_{s}^{2}-2x(1-x)M^{2})}\times\nonumber \\
 &  & \hskip-2cm\bigg[\frac{\tilde{b}^{2}}{2}\e^{\tilde{b}\tilde{m}_{\Lambda}^{2}}(\Gamma(0,\tilde{b}\tilde{m}_{\Lambda}^{2})-\Gamma(0,\tilde{b}(\vec{p}_{T}^{2}+\tilde{m}_{\Lambda}^{2}))+\frac{1-\tilde{b}\tilde{m}_{\Lambda}^{2}}{2\tilde{m}_{\Lambda}^{4}}-\frac{1-\tilde{b}\left(\vec{p}_{T}^{2}+\tilde{m}_{\Lambda}^{2}\right)}{2\left(\vec{p}_{T}^{2}+\tilde{m}_{\Lambda}^{2}\right)^{2}}\e^{-\tilde{b}\vec{p}_{T}^{2}}\bigg].\label{eq:BMsc}\end{eqnarray}

\begin{figure}[top]
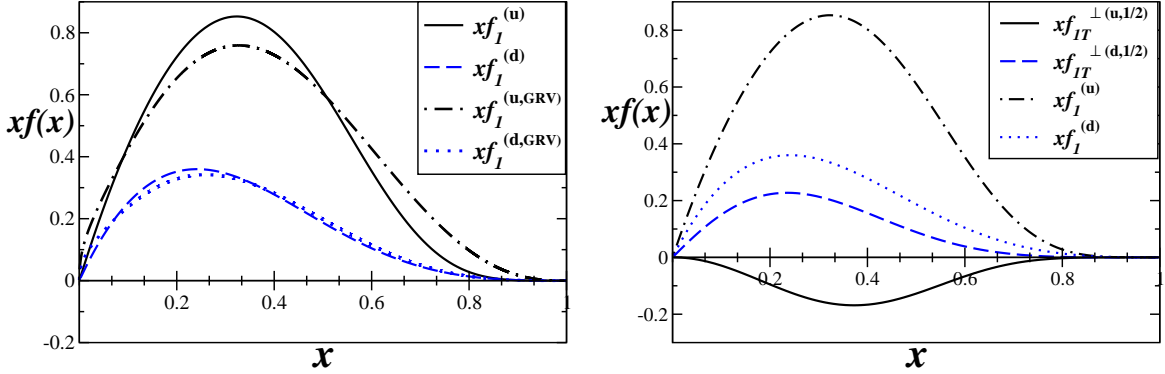

\begin{center}
\includegraphics[scale=0.3]{un_axialpubgrvnewLOwk.eps}~
~\includegraphics[scale=0.3]{siv_axialpubnew.eps}
\caption{\small{Left Panel: The unpolarized $u$- and $d$-quark distributions 
functions versus $x$ compared to
the low scale parameterization of the unpolarized $u$- and $d$- 
quark distributions~\cite{Gluck:1998xa}.  Right Panel: The half moment of 
the  Sivers functions 
and the unpolarized $u$ and $d$ distributions  versus $x$ compared to
the low scale parameterization of the unpolarized $u$- and $d$- 
quark distributions ($\kappa=1.0$).
}\label{cap:xfg}}
\end{center}
\end{figure}
\subsection{Sivers-function in the diquark spectator model}
\begin{figure}[bottom]
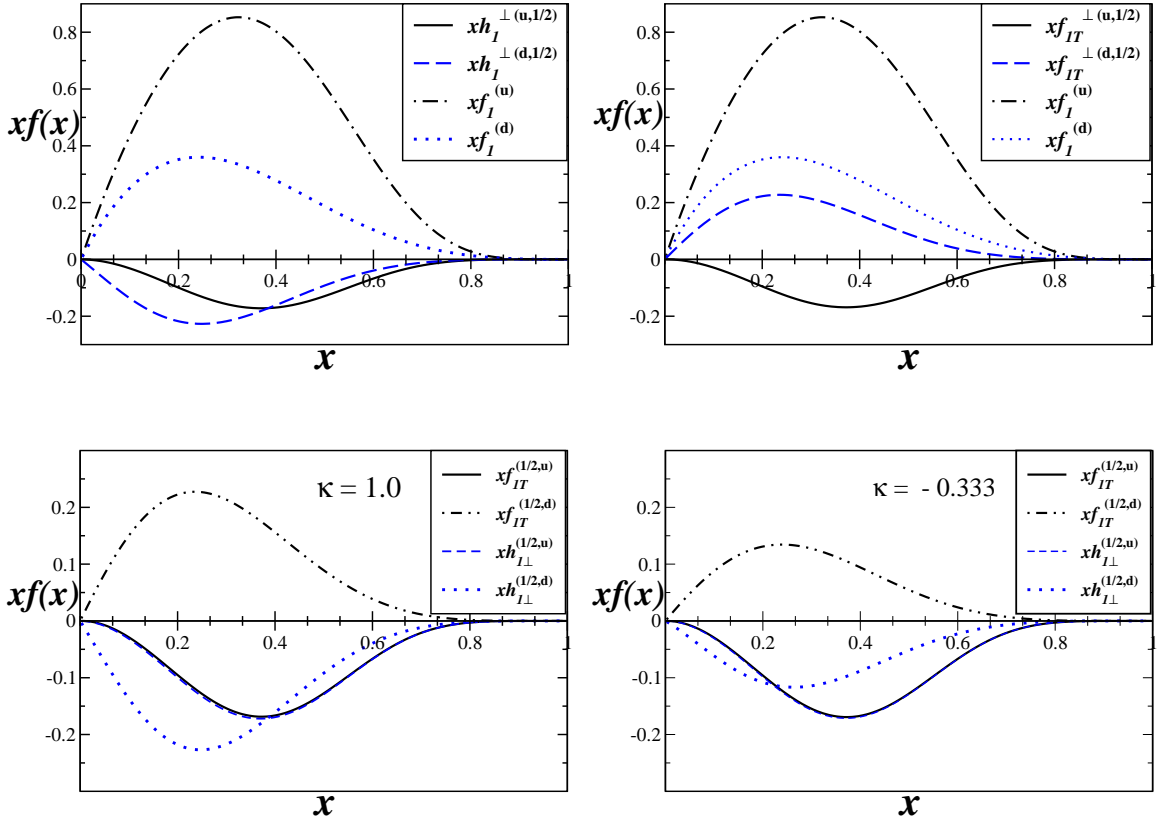

\begin{center}
\includegraphics[scale=0.3]{bm_axialpubnew.eps}~
\includegraphics[scale=0.3]{siv_axialpubnew.eps}\vskip 1.cm
\includegraphics[scale=0.3]
{sivbm1_axialpubnew.eps}~
\includegraphics[scale=0.3]
{sivbm3_axialpubnew.eps}
\caption{\small{Top Panels: The half-moment of the Boer Mulders (left) 
and Sivers (right)
functions versus $x$ compared to the unpolarized $u$- and $d$-  quark
distribution functions.  Bottom Panels: The half moments of the Boer  Mulders and Sivers functions, $\kappa=1.0$ lower-left, $\kappa=-0.333$ lower-right, 
versus $x$, extractions from data were presented in 
Ref.~\cite{Anselmino:2005an,Anselmino:2005ea,Vogelsang:2005cs,
Collins:2005ie,Collins:2005rq} 
for the Sivers function.} \label{cap:xf}}
\end{center}
\end{figure}
Having obtained the results for the Boer-Mulders function $h_{1}^{\perp}$,
it is straight-forward to apply the procedure described above to calculate
the Sivers-function $f_{1T}^{\perp}$. The Sivers-function can be
extracted from the following trace of the quark-quark correlator (\ref{eq:Correlator})
(see e.g. \cite{Goeke:2005hb,Bacchetta:2006tn}),\begin{equation}
2S_{T}^{i}\epsilon_{T}^{ij}p_{T}^{j}f_{1T}^{\perp}(x,\vec{p}_{T}^{2})=\frac{M}{2}\int dp^{-}\,\left(\Tr\left[\gamma^{+}\Phi(p;P,S_{T})\right]-\Tr\left[\gamma^{+}\Phi(p;P,-S_{T})\right]\right)\bigg|_{p^{+}=xP^{+}}.\end{equation}

It is well-known \cite{Goldstein:2002vv} that in the scalar diquark
spectator sector the Boer-Mulders function and the Sivers function
coincide, so the scalar Sivers function is given by the left-hand
side of Eq.~(\ref{eq:BMsc}). By contrast the different Dirac structure
for the chiral even Sivers function and chiral odd Boer-Mulders function
in the axial-vector diquark sector, Eq. (\ref{bm}) and (\ref{eq:StartingFormula})
respectively, lead to different coefficients in the decomposition
$N_{\alpha_{1}...\alpha_{i}}^{(i)}$ in Eq. (\ref{eq:NumDecomp}).
So, whereas the  form of the Sivers-function $f_{1T}^{\perp,ax}$ is the same as
the form of $h_{1}^{\perp,ax}$ given in Eq. (\ref{eq:BMax}), 
the coefficients $A_{p}$, $B_{p}$, $C_{p}$, $D_{p}$, $E_{p}$,
$A_{l}$, $B_{l}$, $C_{l}$, $D_{l}$, $E_{l}$, $A_{lp}$, $B_{lp}$,
$E_{lp}$ differ. They are
given for $f_{1T}^{\perp,ax}$ explicitly in the Appendix \ref{append2}.
\begin{figure}[bottom]
\begin{center}
\includegraphics[scale=0.3]{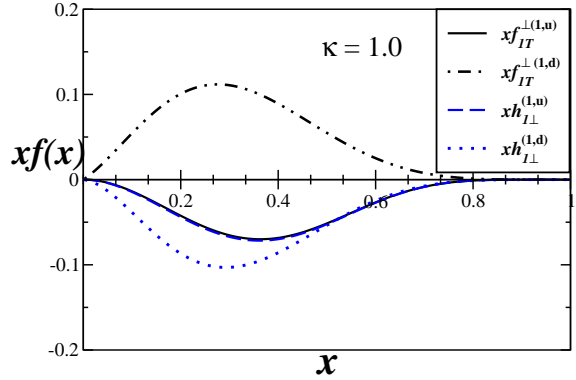}
\caption{\small{
 The first  moment of the Boer-Mulders and Sivers functions 
 versus $x$ for  $\kappa=1.0$.
}\label{cap:first}}
\end{center}
\end{figure}

\section{Asymmetries in SIDIS}

\subsection{Azimuthal $\cos(2\phi)$-asymmetry in unpolarized SIDIS}

\begin{figure}
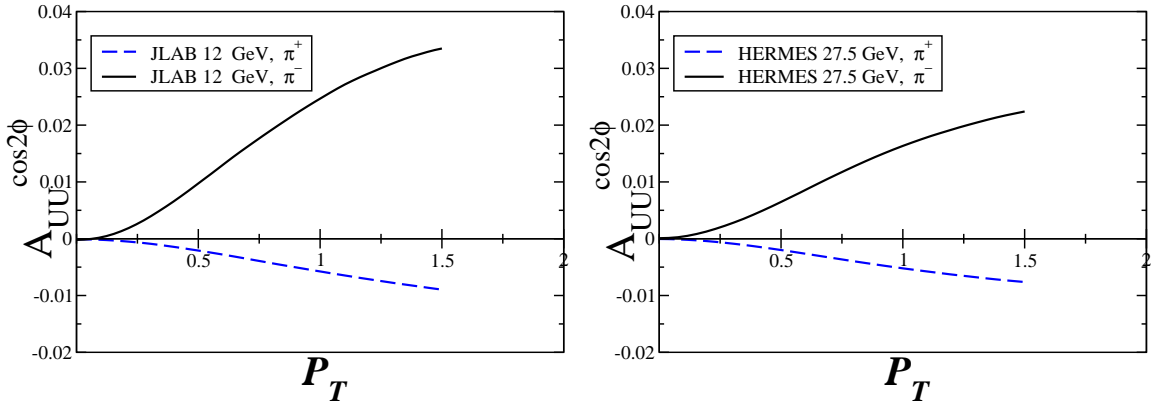

\begin{center}
\includegraphics[scale=0.3]{cospt_pubnew.eps}
~\includegraphics[scale=0.3]{cospt_pubhnew.eps}
\caption{\small{Left Panel: The $\cos2\phi$ asymmetry for $\pi^{+}$
and $\pi^{-}$ as a function of $P_{T}$ at JLAB $12\textrm{GeV}$
kinematics. Right Panel: The $\cos2\phi$ asymmetry for $\pi^{+}$ and
$\pi^{-}$ as a function $P_{T}$ for HERMES kinematics.} 
\label{asym:pt}}
\end{center}
\end{figure}

\begin{figure}
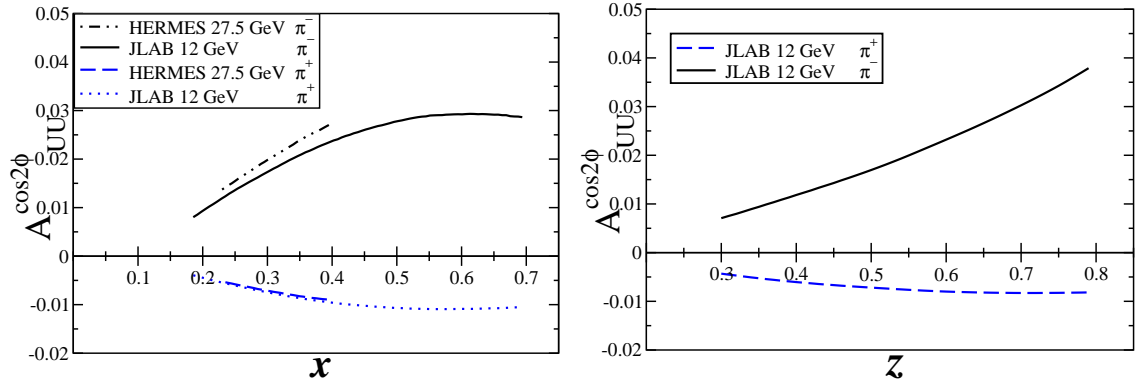

\begin{center}
\includegraphics[scale=0.3]{cosx_pubnew.eps}
~\includegraphics[scale=0.3]{cosz_pubnew.eps}
\caption{\small{Left Panel: The $\cos2\phi$ asymmetry for $\pi^{+}$ and
$\pi^{-}$ as a function of $x$ at JLAB $12\textrm{GeV}$ 
and HERMES kinematics. Right Panel: The $\cos2\phi$ asymmetry for $\pi^{+}$ and
$\pi^{-}$ as a function $z$ for JLAB 
kinematics.} \label{figxz}}
\end{center}
\end{figure}

Almost 30 years ago  it was pointed out that both kinematic
\cite{Cahn:1978se} and dynamical effects \cite{Berger:1979kz,Berger:1979xz}
could give rise to an $\cos2\phi$ azimuthal asymmetry going like
$p_{T}^{2}/Q^{2}$ (where $Q$ is a hard scale) 
 when transverse momentum scales are on the order
of the intrinsic momentum scales of partons, $P_{T}\sim p_{T}$. However,
when transverse momentum is on the order of the hard scale, $P_{T}\sim Q$,
these non-perturbative effects are expected to decrease relative to
perturbative contributions \cite{Georgi:1977tv,Mendez:1978zx}. By
contrast, taking into account the existence of ``T-odd'' TMDs and
fragmentation functions it was pointed out by Boer and Mulders~\cite{Boer:1997nt}
that at leading twist a convolution of the Boer-Mulders and the Collins
functions would give rise to non-trivial azimuthal asymmetries in
unpolarized SIDIS.

Having explored the flavor dependence of the $h_{1}^{\perp}$ we are
now in a position to extend early phenomenological work on ``T-odd'' contributions
to azimuthal asymmetries in SIDIS performed under the approximation
of scalar diquark dominance\cite{Gamberg:2003ey}. In particular we
consider the spin independent double ``T-odd'' $\cos2\phi$ asymmetry
for $\pi^{+}$ and $\pi^{-}$ production.

The general form of the cross section for an unpolarized target  reads
\cite{Bacchetta:2006tn}\begin{eqnarray}
\frac{d\sigma}{dx\, dy\,\, dz\, d\phi_{h}\, dP_{h\perp}^{2}} & \approx & \frac{2\pi\alpha^{2}}{xyQ^{2}}\,\Big[\left(1-y+\frac{1}{2}y^{2}\right)\, F_{UU,T}\,+\,\left(1-y\right)\, F_{UU,L}\\
 &  & +\left(2-y\right)\cos(\phi_{h})\,
		       F_{UU}^{\cos\phi}\,+\,\left(1-y\right)\cos(2\phi_{h})\, F_{UU}^{\cos2\phi_{h}}\Big],\nonumber \end{eqnarray} 
where the structure function $F_{UU}^{\cos2\phi_{h}}$ is of most
interest for the purpose of this paper. At leading twist it factorizes into a convolution
of the Boer-Mulders and Collins fragmentation function \cite{Ji:2004wu,Bacchetta:2006tn} \bea
F_{UU}^{\cos 2\phi_h} = \mathcal{C}\biggl[
   - \frac{2  \hat{\bm{h}}\cdot \bm{k}_T\hat{\bm{h}}\cdot \bm{p}_T 
    -\bm{k}_T \cdot \bm{p}_T}{M M_h}
    h_{1}^{\perp } H_{1}^{\perp }\biggr],\label{FUUcos2phi} \eea
the convolution integral $\mathcal{C}$ is given by \begin{equation}
{\mathcal{C}}\bigl[w\, f\, D\bigr]=x\,\sum_{a}e_{a}^{2}\int d^{2}\bm{p}_{T}\, d^{2}\bm{k}_{T}\,\delta^{(2)}\bigl(\bm{p}_{T}-\bm{k}_{T}-\bm{P}_{h\perp}/z\bigr)\, w(\bm{p}_{T},\bm{k}_{T})\, f^{a}(x,p_{T}^{2})\, D^{a}(z,k_{T}^{2}),\label{conv}\end{equation}
 where summation runs over quarks and anti-quarks. $\bm{p}_{T}$, $\bm{k}_{T}$
are the intrinsic transverse momenta of the active and fragmenting
quarks respectively and $\bm{P}_{h}$ is the transverse momentum of
the fragmenting hadron with respect to the photon momentum $q$. $\hat{\bm{h}}$
is defined as $\bm{P}_{h\perp}/|\bm{P}_{h\perp}|$.

We have fixed most of the model parameters such as masses and normalizations
by comparing the model result for the unpolarized ``T-even'' PDF $f_{1}$
for $u$- and $d$- quarks (Eqs. (\ref{f1sc}) and (\ref{f1ax}))  to
the leading order (LO) 
low-scale ($\mu^{2}=0.26{\textrm{GeV}}$) data parameterization
of Gl\"{u}ck, Reya, and Vogt \cite{Gluck:1998xa}. Note that PDFs
for $u$- and $d$- quarks are given by linear combinations of PDFs
for an axial vector and scalar diquark, $u=\frac{3}{2}f^{sc}+\frac{1}{2}f^{ax}$
and $d=f^{ax}$~\cite{Jakob:1997wg,Bacchetta:2003rz}. The best model
approximation to the GRV data parameterization for $u$ and $d$
of \cite{Gluck:1998xa} is shown in Fig. \ref{cap:xfg}, and corresponds
to a set of parameters 
$m_{q}=0\ \rm{GeV}$, $m_{s}=1.0\ \rm {GeV}$, $m_{ax}=1.3\ \rm{GeV}$,
 $\Lambda=1.3\ {\rm GeV}$, $M=0.94\ \rm{ GeV\, -fixed}$, and $R_{g}=5/4$.  
For ``T-odd'' PDFs such a procedure for fixing the model parameters is
not sufficient since it doesn't determine the sign and the strength
of the final state interactions. In our case the final state interactions
are described effectively in the one gluon exchange approximation
by the product $e_{dq}e_{q}$, the charges of the diquark and quark,
respectively. We need to fix the value of this product. For that reason
we calculated the Sivers function for $u$- and $d$- quarks in the
diquark model and compared our results 
in Fig. \ref{cap:xfg}  for the ``one-half'' moment,
\begin{equation}
f_{1T}^{\perp(q,1/2)}(x)=\int{d^2 p_T\,\frac{|\vec{p}_T|}{M}\,f_{1T}^{\perp (q)}(x,\vec{p}_T^2)},
\end{equation}
as well as the first moment,
\begin{equation}
f_{1T}^{\perp(q,1)}(x)=\int{d^2 p_T\,\frac{\vec{p}_T^2}{2M^2}\,f_{1T}^{\perp (q)}(x,\vec{p}_T^2)},
\end{equation}
with the  existing data parameterizations where $q$ represents the quark
flavor (see Refs.~\cite{Anselmino:2005an,Anselmino:2005ea,Vogelsang:2005cs,
Collins:2005ie,Collins:2005rq}). 
In such a way we are able to fix 
$e_{dq}e_{q}/4\pi = C_{F}\alpha_{s}=0.267$~\footnote{This is in
agreement with the value $\alpha_s$ used in \cite{Bacchetta:2007xx} to explore
the ``T-odd'' fragmentation functions. It is worth noting that 
the running coupling 
extrapolated to the   ``low-scale'' $\mu^2$ in~\cite{Gluck:1998xa} 
is different than the coupling that characterizes the FSIs in the
one gluon exchange approximation.}
with color factor, $C_{F}=4/3$. We display the ``one-half'' and first
moments for $u$- and $d$-  quark
Sivers functions $f_{1T}^{\perp(q)}$ and 
Boer-Mulders functions $h_{1}^{\perp(q)}$
(where $q=u,d$) along with the unpolarized $u$- and $d$- quark pdfs
in Figs.~\ref{cap:xf} and ~\ref{cap:first}.  The ``one-half'' 
and first moments of the $u$- and $d$- quark Sivers 
functions are negative and positive respectively
while the $u$- and $d$-  quark Boer Mulders functions are both negative
over the full range in Bjorken-$x$. These results are in agreement
with the large $N_{c}$ predictions \cite{Pobylitsa:2003ty}, Bag
Model results reported in \cite{Yuan:2003wk}, impact parameter distortion
picture of Burkardt \cite{Burkardt:2005hp} and recent studies of
nucleon transverse spin structure in 
lattice QCD~\cite{Gockeler:2006zu}.~\footnote{
It is interesting to note the approximate agreement for the 
flavor dependence of $h_1^\perp$ among such  models 
probably arises because our input quark and
di-quark wave functions share the same SU(4) flavor-spin dependence as the 
bag and other spectator models. Additionally SU(4) 
symmetric baryon wave functions 
are  compatible  with large-$N_c$ counting rules.}
Also, we explored the relative dependence of
the $d$-quark to $u$-quark Sivers function, see Fig.~\ref{cap:xf}. 
For example, choosing a value of $\kappa=-0.333$ as was determined in
Ref.~\cite{Goldstein:1979wb} we find the $d$ quark Sivers is smaller than the
$u$-  quark. Choosing $\kappa=1$ we find reasonable agreement with
extractions reported in \cite{Anselmino:2005an}. 
It is worth noting (see Fig. \ref{cap:xf} )
that the resulting $u$-quark Sivers function and 
Boer-Mulders function are nearly equal, even with the inclusion of the 
axial vector spectator diquark. An exact equality was first noted in the 
simpler scalar di-quark dominance approximation in~\cite{Goldstein:2002vv}.

Our model input for the Collins functions is based on very recent work
in~\cite{Bacchetta:2007xx} where the Collins function was calculated in the spectator framework. Therein it was assumed 
that $H_1^{\perp(dis-fav)}\approx -H_1^{\perp(fav)}$ in the pion sector, thereby satisfying the Schaefer-Teryaev sum rule \cite{Schafer:1999kn} locally.  
We use those results along with the results of this paper for $h_{1}^{\perp}$ to estimate the azimuthal asymmetry $A_{UU}^{\cos2\phi}$ (cf. Eq. (\ref{FUUcos2phi})), where
\begin{equation}
A_{UU}^{\cos 2\phi}\equiv \frac{\int d\Phi \ \cos 2\phi\ d\sigma}{\int d\Phi \ d\sigma}
\end{equation}
and $d\Phi$ is short-hand notation for the phase space integration.
In Fig.~\ref{asym:pt} we display the $A_{UU}^{\cos2\phi}(P_{T})$
in the range of future JLAB kinematics \cite{Clas:2006ka} 
($0.08<x<0.7$, $0.2< y < 0.9$,\linebreak
 $0.3< z < 0.8$, $Q^2 > 1 {\textrm{GeV/c}}$,
and $1<E_{\pi}<9 \ {\textrm{GeV}}$) 
and HERMES kinematics \cite{Airapetian:2004tw}
 ($0.23<x<0.4$,
$0.1 <y<0.85$, $0.2<z<0.7$, with $Q^2 > 1\ {\textrm{GeV/c}}$,
and $4.5 <E_{\pi}<13.5\ {\textrm{GeV}}$).  In Fig.~\ref{figxz}
we display the $x$ and $z$ dependence in the 
range $0.5 <P_{T}<1.5\,\, \rm{GeV/c}$. It should be noted that this asymmetry was measured at HERA by ZEUS, but at very low $x$ and very high $Q^2$~ \cite{Chekanov:2006gt} where other QCD effects dominate. It was also measured at CERN by EMC~\cite{Arneodo:1986cf}, but with low precision. Those data were approximated by Barone, Lu and Ma~\cite{Barone:2005kt} in a $u$-quark dominating model for $h_1^\perp$, with a Gaussian, algebraic form and a Gaussian ansatz for the Collins function. Our dynamical approach leads 
to different predictions for the forthcoming JLab data.

\subsection{Single spin asymmetry $A_{UL}^{\sin(2\phi)}$ in SIDIS}

\begin{figure}[top]
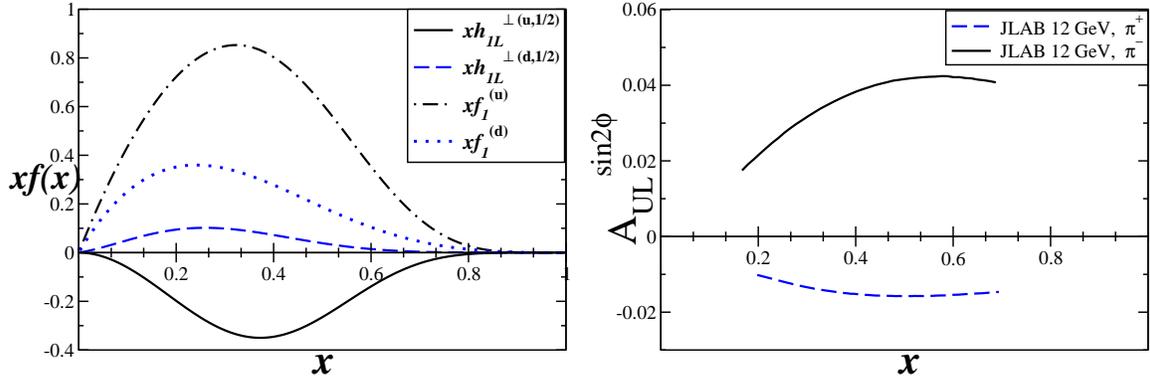

\begin{center}
\includegraphics[scale=0.3]{h1Lnew.eps}
~\includegraphics[scale=0.3]{sinx_pubnew.eps}
\caption{\small{Left Panel:
The half-moment of $xh_{1L}^{\perp(1/2)}$
 versus $x$ compared to the unpolarized $u$- and $d$- quark
distribution functions. Right Panel: The $\sin2\phi$ asymmetry for 
$\pi^{+}$ and $\pi^{-}$ as a function of $x$ at JLAB $12\textrm{GeV}$ 
kinematics.}}
\label{h1L}
\end{center}
\end{figure}

Since we have calculated the chiral-odd but ``T-even'' parton distribution
$h_{1L}^{\perp}$  (cf. Eqs. (\ref{h1Lperpax}),
(\ref{h1Lperpsc})) we use this result together with the result of
Ref. \cite{Bacchetta:2007xx} for the Collins function to give a prediction
for the $\sin(2\phi)$ moment of the single spin asymmetry $A_{UL}$
for a longitudinally polarized target. In particular we are able to
take into account the flavor dependence of the asymmetry. We adopt
the similar procedure for the azimuthal $\cos(2\phi)$-asymmetry for
treating the leading twist observable $A_{UL}^{\sin(2\phi)}$.

A decomposition into structure functions of the cross section of semi-inclusive
DIS for a longitudinally polarized target reads (see e.g. \cite{Bacchetta:2006tn})

\begin{eqnarray}
\frac{d\sigma_{UL}}{dx\, dy\, dz\, d\phi_{h}\, dP_{h\perp}^{2}} & \approx & \frac{2\pi\alpha^{2}}{xyQ^{2}}\, S_{\parallel}\,\Big[(1-y)\sin(2\phi_{h})\, F_{UL}^{\sin(2\phi)}+(2-y)\sqrt{1-y}\sin(\phi_{h})\, F_{UL}^{\sin\phi}\Big],\nonumber \\
\end{eqnarray}
where $S_{\parallel}$ is the projection of the spin vector on the
direction of the virtual photon. In a partonic picture the structure
function $F_{UL}^{\sin(2\phi)}$ is a leading twist object (while
$F_{UL}^{\sin\phi}$ is sub-leading) and given by a convolution of
the TMD $h_{1L}^{\perp}$ and the Collins function (cf. \cite{Bacchetta:2006tn})

\begin{equation}
F_{UL}^{\sin(2\phi)}=\mathcal{C}\Bigg[-\frac{2\hat{\bm{h}}\cdot\bm{k}_{T}\,\hat{\bm{h}}\cdot\bm{p}_{T}-\bm{k}_{T}\cdot\bm{p}_{T}}{MM_{h}}\, h_{1L}^{\perp}H_{1}^{\perp}\Bigg],\label{FULsin2phi}\end{equation}
 where the explicit form of the convolution is given in Eq. (\ref{conv}).

We insert our result for $h_{1L}^{\perp}$ (Eqs. (\ref{h1Lperpax}),
(\ref{h1Lperpsc})) and the result of Ref. \cite{Bacchetta:2007xx}
into Eq. (\ref{FULsin2phi}) to compute the single spin asymmetry.
This is the first calculation of this observable in the spectator
framework, whereas the part of $F_{UL}^{\sin(\phi)}$ described by
higher twist ``T-odd'' PDFs has been analyzed in the diquark model in
Refs. \cite{Afanasev:2003ze,Metz:2004je,Afanasev:2006gw}. Similar
phenomenology for $F_{UL}^{\sin(2\phi)}$ and $F_{UL}^{\sin(\phi)}$
has been performed in Refs. \cite{Efremov:2001cz,Efremov:2002ut}
using the framework of the chiral quark soliton model.

We display the results for the single spin asymmetry $A_{UL}^{\sin(2\phi)}$
in Fig.~\ref{h1L} using the kinematics of the upcoming JLab 12 GeV upgrade. We note that the $\pi^-$ asymmetry is large and positive due to the model assumption $H_1^{\perp(dis-fav)}\approx -H_1^{\perp(fav)}$. This asymmetry has been measured at HERMES for longitudinally polarized
protons~\cite{Airapetian:1999tv} and deuterons~\cite{Airapetian:2002mf}. The data show that for the proton target and HERMES 27.5 GeV kinematics both $\pi^+$ and $\pi^-$ asymmetries are consistent with 0 down to a sensitivity of about 0.01. That is to say, these asymmetries could be non-zero, but with magnitudes less than 0.01 or 0.02. These results are considerably smaller than our predictions for the JLab upgrade. For the deuteron target the results are  consistent with 0 for $\pi^+$ and $\pi^-$. There is one $\pi^0$ point at $x\sim 0.2$ that could be positive at about 0.03. This SIDIS data for polarized deuterons could reflect the near cancellation of $u$- and $d$-quark $h_{1L}^{\perp}$ functions and/or the large unfavored Collins function contributions. There is also CLAS preliminary data~\cite{Avakian:2005ps} at 5.7 GeV that shows slightly negative asymmetries for $\pi^+$ and $\pi^-$ and leads to the extraction of a negative $xh_{1L}^{\perp(u)}$. This suggests that the unfavored Collins function (for $d\rightarrow\pi^+$) is not contributing much here, unlike the inference from the HERMES data. Data from the upgrade should help resolve these phenomenological questions.

\section{Conclusions}

In this paper we performed calculations of transverse momentum
dependent parton distributions, including the Boer-Mulders function 
$h_{1}^{\perp}$, the Sivers function $f_{1T}^{\perp}$,  
and $h_{1L}^{\perp}$ in the diquark spectator model taking into account 
both axial-vector and scalar contributions.
The calculation of these functions in both
sectors allowed us to explore their flavor dependence, i.e. to compute 
their  $u$-quark and a $d$-quark contributions. For ``T-even''
 distributions like $h_{1L}^{\perp}$, a non-trivial contribution
could already have been obtained from a tree-level diagram. 
By contrast,  final state interactions or, equivalently, contributions from
the gauge link had to be taken into account for the ``T-odd'' 
 Boer-Mulders  and  Sivers functions, requiring the calculation of a loop (box) diagram. It was found that
the loop integrals for $f_{1T}^{\perp}$ and $h_{1}^{\perp}$ show
light cone divergences and UV divergences in the axial-vector diquark
sector, in contrast to the scalar diquark sector. We regularized these
divergences by choosing specific types of the phenomenologically motivated
nucleon-diquark-quark form factors. By comparing the model expression
 for the unpolarized parton  distribution, $f_{1}$, 
with the low scale parameterizations of that function
obtained from data, it is possible to fix most of the parameters of
the model, masses, normalization and $R_g$, the ratio of axial diquark couplings to the nucleon.
In order to fix the sign and size of the final state interactions
specific for ``T-odd'' distributions, we calculated the Sivers function
and compared the result to parameterizations of SIDIS data already determined in a global fit 
for $f_{1T}^{\perp}$. In such a way the remaining parameters
could be fixed, and  predictions were presented for the flavor dependence of
the Boer-Mulders
function $h_{1}^{\perp}$ for a $u$-quark and a $d$-quark. 
We find that the $k_T$-half- and first-moments  
of  this function are negative  for both flavors.  
This result is in  
contrast to that in~\cite{Bacchetta:2003rz}. Our sign result 
is in agreement with   other approaches that 
predict negative $h_{1}^{\perp(u)}$ and $h_{1}^{\perp(d)}$.

We also note that the resulting $u$-quark Sivers function and 
Boer-Mulders function are nearly equal, even with the inclusion of the 
axial vector spectator diquark. This near equality $h_1^\perp \sim 
f_{1T}^\perp$ was obtained from models without axial di-quarks, hinting 
at some more general mechanism that preserves the relation.

We used our result for $h_{1}^{\perp}$ as one ingredient in the
factorized formula for the azimuthal asymmetry $A_{UU}^{(\cos(2\phi))}$
in unpolarized semi-inclusive lepto-production of positively and negatively
charged pions. We also used our $h_{1L}^{\perp}$ as an ingredient in the
single-spin asymmetry $A_{UL}^{(\sin(2\phi))}$ for a longitudinally
polarized target in semi-inclusive DIS. Another key ingredient for determining
such asymmetries is the Collins fragmentation function $H_{1}^{\perp}$.
For this function we used the most current expressions that were obtained in a similar spectator model. 

We provide estimates for $A_{UU}^{(\cos(2\phi))}$ and $A_{UL}^{(\sin(2\phi))}$.
The latter has already been measured at HERMES and preliminarily by CLAS. There are important differences in the kinematic regions explored, but there remain discrepancies that may be resolved in the future at Jefferson Lab, for which our model gives striking predictions of relatively large asymmetries. The 
non-trivial $\pi^-$ asymmetry is  driven in large part by
the model assumption $H_1^{\perp(dis-fav)}\approx -H_1^{\perp(fav)}$ in the pion sector.  We note that our result for $A_{UL}^{(\sin(2\phi))}$ is the first phenomenological
treatment in the spectator framework of this observable. 

The former unpolarized 
asymmetry, $A_{UU}^{(\cos(2\phi))}$, was measured at HERA, but for very small $x$ and extremely high $Q^2$. Again, this will be measured in the future at JLab. We predict that those results should correspond to the opposite sign  asymmetries for opposite charged pions.

In summary, a refined diquark spectator model, including axial vector di-quarks leads to 
both $u$- and $d$-quark ``T-odd'' TMDs and provides the ingredients for predicting a 
range of asymmetries for future experiments.   The approach we have been taking is to use and refine a model for the 
soft regime that makes sense in QCD and can be applied broadly to a 
range of measurable phenomena.  We have fixed the parameters in the 
model to approach the inferred structure of the lowest order 
asymmetries. Combined with the recent determination of the fragmentation 
functions, we have predicted new SIDIS results. The spirit of this work 
is to understand the dynamics of processes like SIDIS by refining a 
robust and flexible model for the ``T-odd'' functions that compares with 
existing data. While a global fit to all the data eventually can be 
performed, the underlying mechanism is likely to be revealed by honing 
in on more sophisticated and inclusive models, as we have done here.
\\

\textbf{\emph{Acknowledgments:}} We thank Alessandro Bacchetta and
Asmita Mukherjee for fruitful discussions and use of results from
hep-ph/0707.3372.\\

\textbf{Notice:} Authored by Jefferson Science Associates, LLC under
U.S. DOE Contract No. DE-AC05-06OR23177. The U.S. Government retains
a non-exclusive, paid-up, irrevocable, world-wide license to publish
or reproduce this manuscript for U.S. Government purposes. This work
is supported in part by the U.S. Department of Energy under contracts,
DE-FG02-07ER41460 (LG), and DE-FG02-92ER40702 (GRG).
\vspace{.5cm}

\appendix

\section{Appendix: Axial Diquark Contribution to Boer Mulders Function\label{append1}}

The coefficients appearing in the calculation of the Boer Mulders
function for the axial vector diquark (see text) read

{\small \begin{eqnarray}
A_{p} & = & (1-x)\Big[(2Mm_{q}+(2-R_{g})M^{2}+R_{g}m_{q}^{2})(3+2\kappa)+R_{g}m_{s}^{2}(7+4\kappa)\Big],\end{eqnarray}
\begin{eqnarray}
B_{p} & = & (1-x)\Big[(2Mm_{q}+(2-R_{g})M^{2}+R_{g}m_{q}^{2})(3+2\kappa)+R_{g}m_{s}^{2}(11+6\kappa)\Big],\end{eqnarray}
\begin{eqnarray}
C_{p} & = & 2m_{s}^{2}(1-x)\Big[2Mm_{q}(-1+2x(2+\kappa))+R_{g}\big(m_{s}^{2}+6\vec{p}_{T}^{2}(2+\kappa)+m_{q}^{2}(3+2\kappa)\big)\nonumber \\
 &  & +M^{2}\left(-2+4x(2+\kappa)+R_{g}(1-2x(2-x)(2+\kappa))\right)\Big],\end{eqnarray}
\begin{eqnarray}
D_{p} & = & 2m_{s}^{2}(1-x)\Big[2Mm_{q}(-1+2x(2+\kappa))+R_{g}\big(m_{s}^{2}+2\vec{p}_{T}^{2}(2+\kappa)+m_{q}^{2}(3+2\kappa)\big)\nonumber \\
 &  & +M^{2}\left(-2+4x(2+\kappa)+R_{g}(1-2x(2-x)(2+\kappa))\right)\Big],\end{eqnarray}
\begin{eqnarray}
E_{p} & = & 0,\end{eqnarray}
\begin{eqnarray}
A_{l} & = & -\frac{R_{g}(3+2\kappa)}{2M}\Big[R_{g}m_{q}(m_{s}^{2}-\vec{p}_{T}^{2})+M^{2}m_{q}(2-R_{g})(1-x)^{2}+x(1-x)^{2}R_{g}M^{3}\nonumber \\
 &  & +M\left((2-R_{g})xm_{s}^{2}-\vec{p}_{T}^{2}(2-xR_{g})\right)\Big],\end{eqnarray}
\begin{eqnarray}
B_{l} & = & -\frac{R_{g}}{2M}\Big[(3+2\kappa)\big[M^{2}m_{q}(2-R_{g})(1-x)^{2}+x(1-x)^{2}R_{g}M^{3}\nonumber \\
 &  & +m_{q}R_{g}\left(-\vec{p}_{T}^{2}+m_{s}^{2}\frac{7+4\kappa}{3+2\kappa}\right)\big]\nonumber \\
 &  & +M\left[-\vec{p}_{T}^{2}(2-xR_{g})(3+2\kappa)+m_{s}^{2}\left(4(2+\kappa)+x(6-7R_{g}+4\kappa(1-R_{g}))\right)\right]\Big],\end{eqnarray}
\begin{eqnarray}
C_{l} & = & -\frac{1}{2M}\Bigg[(1-x)^{3}(3+2\kappa)M^{4}m_{q}\left(-4+(4-R_{g})(1-x)R_{g}\right)\nonumber \\
 &  & -(1-x)^{3}x(3+2\kappa)M^{5}R_{g}(2-(1-x)R_{g})\nonumber \\
 &  & +m_{q}R_{g}^{2}\left(8m_{s}^{2}\vec{p}_{T}^{2}(2+\kappa)+(m_{s}^{2}-\vec{p}_{T}^{2})(m_{s}^{2}+\vec{p}_{T}^{2})(3+2\kappa)\right)\nonumber \\
 &  & +MR_{g}\Big(xm_{s}^{4}(2-R_{g})(3+2\kappa)-\vec{p}_{T}^{4}(2-xR_{g})(3+2\kappa)\nonumber \\
 &  & -2m_{s}^{2}\vec{p}_{T}^{2}\left(-17+x+8xR_{g}-8\kappa+4xR_{g}\kappa\right)+2m_{q}^{2}(1-x)(m_{s}^{2}+(3+2\kappa)\vec{p}_{T}^{2})\Big)\nonumber \\
 &  & +2(1-x)M^{2}m_{q}\Big(\vec{p}_{T}^{2}(3+2\kappa)\left(2+(1-x)R_{g}(2-R_{g})\right)\nonumber \\
 &  & +m_{s}^{2}(4(2+\kappa)-x(6+4\kappa))\Big)
+2(1-x)M^{3}\Big(-m_{q}^{2}(2-R_{g})(1-x)^{2}(3+2\kappa)\nonumber \\
 &  & +\vec{p}_{T}^{2}\left(2-R_{g}+x(1-x)R_{g}^{2}\right)(3+2\kappa)+xm_{s}^{2}\left(2+R_{g}(6-7x+4\kappa(1-x)\right)\Big)\Big)\Bigg],\end{eqnarray}
\begin{eqnarray}
D_{l} & = & \frac{m_{s}^{2}}{M}\Bigg[-m_{q}R_{g}^{2}(m_{s}^{2}+\vec{p}_{T}^{2})+(1-x)^{2}M^{2}m_{q}R_{g}(2-R_{g})\nonumber \\
 &  & +M^{3}(1-x)^{2}\left(xR_{g}^{2}+4(2+\kappa)-2R_{g}(1+x)(2+\kappa)\right)\nonumber \\
 &  & +MR_{g}\Big(m_{s}^{2}(2\kappa-x(2-R_{g}+2\kappa))+\vec{p}_{T}^{2}(x(4+R_{g}+2\kappa)-2(3+\kappa))\Big)\Bigg],\end{eqnarray}
\begin{eqnarray}
E_{l} & = & \frac{m_{s}^{2}}{M}\Bigg[-m_{q}R_{g}^{2}(\vec{p}_{T}^{2}+m_{s}^{2})^{2}+M^{4}m_{q}(1-x)^{3}\Big(-4+(1-x)(4-R_{g})R_{g}\nonumber \\
 &  & +8x(2+\kappa)\Big)-x(1-x)^{3}M^{5}R_{g}\left(2-(1-x)R_{g}-4x(2+\kappa)\right)\nonumber \\
 &  & -2M^{2}m_{q}(1-x)\Big(m_{s}^{2}\left(4+(1-x)(2-R_{g})R_{g}-2x(3+2\kappa)\right)\nonumber \\
 &  & +\vec{p}_{T}^{2}\left(-2-(1-x)(2-R_{g})R_{g})+4x(2+\kappa)\right)\Big)\nonumber \\
 &  & +MR_{g}\Big(-x(2-R_{g})m_{s}^{4}+2m_{s}^{2}\vec{p}_{T}^{2}\left(2\kappa-1-x(1-R_{g}+2\kappa)\right)\nonumber \\
 &  & +2m_{q}^{2}(1-x)\left(m_{s}^{2}(1+2\kappa)-\vec{p}_{T}^{2}(3+2\kappa)\right)+\vec{p}_{T}^{4}\left(x(8+R_{g}+4\kappa)-2(5+2\kappa)\right)\Big)\nonumber \\
 &  & -2(1-x)M^{3}\Big(-m_{q}^{2}(1-x)^{2}(2-R_{g})(3+2\kappa)\nonumber \\
 &  & +xm_{s}^{2}\left(6-8x+R_{g}(-2+R_{g}+3x-xR_{g})-2x\kappa(2-R_{g})\right)\nonumber \\
 &  & +\vec{p}_{T}^{2}\Big(-x(1-x)R_{g}^{2}+R_{g}\left(5+2\kappa-4x(2+\kappa)\right)+2(-5-2\kappa+4x(2+\kappa))\Big)\Big)\Bigg],\end{eqnarray}
\begin{eqnarray}
A_{lp}=B_{lp} & = & -4R_{g}m_{s}^{2}(1-x)(2+\kappa),\end{eqnarray}
\begin{eqnarray}
E_{lp} & = & 4(1-x)m_{s}^{2}\Big(2Mm_{q}(1-x)(2+\kappa)+M^{2}(1-x)(2-R_{g}(1-x))(2+\kappa)\nonumber \\
 &  & +R_{g}\left(\kappa m_{s}^{2}-\vec{p}_{T}^{2}(2+\kappa)\right)\Big).\end{eqnarray}
}{\small \par}

\section{Axial Diquark Contribution to Sivers Function}

\label{append2} For the Sivers function the corresponding coefficients
read{\small \begin{eqnarray}
A_{p}^{\mathrm{Siv.}} & = & -(1-x)(3+2\kappa)\left(2Mm_{q}+(2-R_{g})M^{2}+R_{g}m_{q}^{2}+R_{g}m_{s}^{2}\frac{7+4\kappa}{3+2\kappa}\right),\end{eqnarray}
\begin{eqnarray}
B_{p}^{\mathrm{Siv.}} & = & -(1-x)(3+2\kappa)\left(2Mm_{q}+(2-R_{g})M^{2}+R_{g}m_{q}^{2}+R_{g}m_{s}^{2}\frac{11+6\kappa}{3+2\kappa}\right),\end{eqnarray}
\begin{eqnarray}
C_{p}^{\mathrm{Siv.}} & = & -2(1-x)m_{s}^{2}\Bigg[2Mm_{q}(-1+2x(2+\kappa))+R_{g}\big(m_{s}^{2}+6\vec{p}_{T}^{2}(2+\kappa)+m_{q}^{2}(3+2\kappa)\big)\nonumber \\
 &  & +M^{2}\left(-2+4x(2+\kappa)+R_{g}\left(1-2x(2-x)(2+\kappa)\right)\right)\Bigg],\end{eqnarray}
\begin{eqnarray}
D_{p}^{\mathrm{Siv.}} & = & -2(1-x)m_{s}^{2}\Bigg[2Mm_{q}(-1+2x(2+\kappa))+R_{g}\big(m_{s}^{2}+2\vec{p}_{T}^{2}(2+\kappa)+m_{q}^{2}(3+2\kappa)\big)\nonumber \\
 &  & +M^{2}\left(-2+4x(2+\kappa)+R_{g}\left(1-2x(2-x)(2+\kappa)\right)\right)\Bigg],\end{eqnarray}
\begin{eqnarray}
E_{p}^{\mathrm{Siv.}} & = & 0,\end{eqnarray}
\begin{eqnarray}
A_{l}^{\mathrm{Siv.}} & = & \frac{3+2\kappa}{2M}R_{g}\Bigg[R_{g}m_{q}(m_{s}^{2}-\vec{p}_{T}^{2})+M^{2}m_{q}(2-R_{g})(1-x)^{2}+x(1-x)^{2}R_{g}M^{3}\nonumber \\
 &  & +M\left(xm_{s}^{2}(2-R_{g})-\vec{p}_{T}^{2}(2-xR_{g})\right)\Bigg],\end{eqnarray}
\begin{eqnarray}
B_{l}^{\mathrm{Siv.}} & = & \frac{R_{g}}{2M}\Bigg[(3+2\kappa)\big(M^{2}m_{q}(2-R_{g})(1-x)^{2}+M^{3}R_{g}x(1-x)^{2}\nonumber \\
 &  & +m_{q}R_{g}\left(-\vec{p}_{T}^{2}+m_{s}^{2}\frac{7+4\kappa}{3+2\kappa}\right)-M\vec{p}_{T}^{2}(2-xR_{g})\big)\nonumber \\
 &  & +Mm_{s}^{2}\left(4(2+\kappa)+x\left(6-7R_{g}+4\kappa(1-R_{g})\right)\right)\Bigg],\end{eqnarray}
\begin{eqnarray}
C_{l}^{\mathrm{Siv.}} & = & \frac{1}{2M}\Bigg[-(1-x)^{3}M^{4}m_{q}(4-(4-R_{g})R_{g}(1-x))(3+2\kappa)\nonumber \\
 &  & -x(1-x)^{3}M^{5}R_{g}(2-(1-x)R_{g})(3+2\kappa)\nonumber \\
 &  & +m_{q}R_{g}^{2}\big(8m_{s}^{2}\vec{p}_{T}^{2}(2+\kappa)+(m_{s}^{2}-\vec{p}_{T}^{2})(m_{s}^{2}+\vec{p}_{T}^{2})(3+2\kappa)\big)\nonumber \\
 &  & +MR_{g}\big(xm_{s}^{4}(2-R_{g})(3+2\kappa)-\vec{p}_{T}^{4}(2-xR_{g})(3+2\kappa)\nonumber \\
 &  & +2m_{s}^{2}\vec{p}_{T}^{2}\left(17-x-8xR_{g}+8\kappa-4xR_{g}\kappa\right)+2m_{q}^{2}(1-x)(m_{s}^{2}+\vec{p}_{T}^{2}(3+2\kappa))\big)\nonumber \\
 &  & -2M^{2}m_{q}(1-x)\big(-\vec{p}_{T}^{2}(2+(2-R_{g})R_{g}(1-x))(3+2\kappa)\nonumber \\
 &  & -m_{s}^{2}(4(2+\kappa)-x(6+4\kappa))\big)+2M^{3}(1-x)\big(-m_{q}^{2}(2-R_{g})(1-x)^{2}(3+2\kappa)\nonumber \\
 &  & +\vec{p}_{T}^{2}\left(2-R_{g}+x(1-x)R_{g}^{2}\right)(3+2\kappa)+xm_{s}^{2}\left(2+R_{g}(6-7x+4\kappa(1-x))\right)\big)\Bigg],\end{eqnarray}
\begin{eqnarray}
D_{l}^{\mathrm{Siv.}} & = & -\frac{m_{s}^{2}}{M}\Bigg[-m_{q}R_{g}^{2}(m_{s}^{2}+\vec{p}_{T}^{2})+m_{q}M^{2}(1-x)^{2}R_{g}(2-R_{g})\nonumber \\
 &  & +(1-x)^{2}M^{3}\left(xR_{g}^{2}+4(2+\kappa)-2R_{g}(1+x)(2+\kappa)\right)\nonumber \\
 &  & +MR_{g}\left(m_{s}^{2}(x(-2+R_{g}-2\kappa)+2\kappa)+\vec{p}_{T}^{2}(-2(3+\kappa)+x(4+R_{g}+2\kappa))\right)\Bigg],\end{eqnarray}
\begin{eqnarray}
E_{l}^{\mathrm{Siv.}} & = & -\frac{m_{s}^{2}}{M}\Bigg[-m_{q}\left(m_{s}^{2}+\vec{p}_{T}^{2}\right)^{2}R_{g}^{2}\nonumber \\
 &  & +M^{4}m_{q}(1-x)^{3}\left(-4+(4-R_{g})R_{g}(1-x)+8x(2+\kappa)\right)\nonumber \\
 &  & -M^{5}R_{g}x(1-x)^{3}\left(2-(1-x)R_{g}-4x(2+\kappa)\right)\nonumber \\
 &  & +2M^{2}m_{q}(1-x)\Big(\vec{p}_{T}^{2}(2+(2-R_{g})R_{g}(1-x)-4x(2+\kappa))\nonumber \\
 &  & +m_{s}^{2}(2x+4x\kappa-(2-R_{g})R_{g}(1-x))\Big)\nonumber \\
 &  & +MR_{r}\Big(-xm_{s}^{4}(2-R_{g})+2m_{s}^{2}\vec{p}_{T}^{2}\left(2\kappa-1-x(1-R_{g}+2\kappa)\right)\nonumber \\
 &  & +2m_{q}^{2}(1-x)\left(m_{s}^{2}(1+2\kappa)-\vec{p}_{T}^{2}(3+2\kappa)\right)\nonumber \\
 &  & +\vec{p}_{T}^{4}(-2(5+2\kappa)+x(8+R_{g}+4\kappa))\Big)\nonumber \\
 &  & -2M^{3}(1-x)\Big(-m_{q}^{2}(2-R_{g})(1-x)^{2}(3+2\kappa)\nonumber \\
 &  & +xm_{s}^{2}\left(2-4x-R_{g}(2-R_{g}-3x+xR_{g})-2(2-R_{g})x\kappa\right)\nonumber \\
 &  & +\vec{p}_{T}^{2}\left(-x(1-x)R_{g}^{2}+R_{g}(5+2\kappa-4x(2+\kappa))-2(5+2\kappa-4x(2+\kappa)\right)\Big)\Bigg],\end{eqnarray}
\begin{eqnarray}
A_{lk}^{\mathrm{Siv.}}=B_{lk}^{\mathrm{Siv.}}=E_{lk}^{\mathrm{Siv.}} & = & 0.\end{eqnarray}
}{\small \par}

\section{$\mathcal{{R}}$-functions\label{append3}}

The $\mathcal{R}$-functions appearing in the text are defined in
the following. The $\mathcal{R}$-function for the unpolarized PDF
$f_{1}$ for an axial vector diquark reads{\small \begin{eqnarray}
&  & \hskip-2cm\mathcal{R}_{1}^{ax}\left(x,\vec{p_{T}^{2}};R_{g},\{\mathcal{M}\}\right)\\ & = & M^{2}\bigg[(\vec{p}_{T}^{2})^{2}+2(1-x(1-x))m_{s}^{2}\vec{p}_{T}^{2}+x^{2}m_{s}^{4}+6x(1-x)^{2}m_{q}Mm_{s}^{2}\nonumber \\
 &  & +(1-x)^{2}M^{2}(\vec{p}_{T}^{2}+2x^{2}m_{s}^{2}+(1-x)^{2}m_{q}^{2})+m_{q}^{2}(1-x)^{2}(\vec{p}_{T}^{2}+2m_{s}^{2})\bigg]\nonumber \\
 &  & +R_{g}\bigg[M^{2}\Big(-(\vec{p}_{T}^{2})^{2}-x^{2}m_{s}^{4}+(1-(4-x)x)m_{s}^{2}\vec{p}_{T}^{2}\nonumber \\
 &  & -(1-x)^{2}m_{q}^{2}(\vec{p}_{T}^{2}-m_{s}^{2})-M^{2}(1-x)^{2}(\vec{p}_{T}^{2}-x^{2}m_{s}^{2}+(1-x)^{2}m_{q}^{2}\Big)\nonumber \\
 &  & +m_{q}M\left(x(\vec{p}_{T}^{2}+m_{s}^{2})^{2}+2x(1-x)^{2}M^{2}(\vec{p}_{T}^{2}-m_{s}^{2}+x(1-x)^{4}M^{2})\right)\bigg]\nonumber \\
 &  & +R_{g}^{2}\bigg[\frac{1}{4}\left(\vec{p}_{T}^{2}+(m_{s}-(1-x)M)^{2}\right)\left(\vec{p}_{T}^{2}+(xM-m_{q})^{2}\right)\nonumber \\
 &  & \left(\vec{p}_{T}^{2}+(m_{s}+(1-x)M)^{2}\right)\bigg].\end{eqnarray}
}The corresponding $\mathcal{R}$-function for $h_{1L}^{\perp,ax}$
reads{\small \begin{eqnarray}
&  & \hskip-2cm\mathcal{R}_{1L}^{\perp,ax}\left(x,\vec{p_{T}^{2}};R_{g},\{\mathcal{M}\}\right)\\ & = & -m_{q}R_{g}^{2}(m_{s}^{2}+\vec{p}_{T}^{2})^{2}+2m_{q}M^{2}(1-x)\Big(\vec{p}_{T}^{2}[2+R_{g}(1-x)(2-R_{g})\nonumber \\
 &  & +m_{s}^{2}\left[2(2-x)-R_{g}(1-x)(2-R_{g})\right]\Big)\nonumber \\
 &  & -M^{4}m_{q}(1-x)^{3}\left[4-(4-R_{g})R_{g}(1-x)\right]-x(1-x)^{3}M^{5}R_{g}(2-(1-x)R_{g})\nonumber \\
 &  & +MR_{g}(m_{s}^{2}+\vec{p}_{T}^{2})\left[2(1-x)m_{q}^{2}-xm_{s}^{2}(2-R_{g})-\vec{p}_{T}^{2}(2-xR_{g})\right]\nonumber \\
 &  & +2(1-x)M^{3}\Big(-m_{q}^{2}(1-x)^{2}(2-R_{g})+xm_{s}^{2}(2+R_{g}(2-(1-x)R_{g}-3x))\nonumber \\
 &  & +\vec{p}_{T}^{2}(2-R_{g}+x(1-x)R_{g}^{2}\Big).\end{eqnarray}
}The $\mathcal{R}$-function for the ``T-odd'' functions $h_{1}^{\perp}$
and $f_{1T}^{\perp}$ is more complicated and contains incomplete
Gamma functions (see text) with $a\equiv\tilde{b}\tilde{m}_{\Lambda}^{2}$,
$c\equiv\tilde{b}(\vec{p}_{T}^{2}+\tilde{m}_{\Lambda}^{2})$,
$d\equiv\tilde{b}\vec{p}_{T}^{2}$,
{\small \begin{eqnarray}
&  & \hskip-2cm \mathcal{R}_{1}^{\perp,ax}\left(x,\vec{p}_{T}^{2};R_{g},\kappa,\tilde{b},\Lambda,\{\mathcal{M}\}\right)\\ & = & \mathcal{R}_{1T}^{\perp,ax}\left(x,\vec{p}_{T}^{2};R_{g},\kappa,\tilde{b},\Lambda,\{\mathcal{M}\}\right)\nonumber \\
 & = & (A_{p}-2A_{l}+B_{l})\left[-\tilde{b}\e^{a}(1+\frac{a}{2})\Gamma(0,a)+\frac{\tilde{b}}{2}\frac{1+a}{a}\right]\nonumber \\
 &  & +\left(\vec{p}_{T}^{2}\left[A_{p}-A_{l}-2(B_{p}-B_{l})\right]+C_{p}-C_{l}\right)\left[\frac{\tilde{b}^{2}}{2}\e^{a}\Gamma(0,a)+\tilde{b}^{2}\frac{1-a}{2a^{2}}\right]\nonumber \\
 &  & +\frac{A_{lp}}{2\vec{p}_{T}^{2}}\left[(1+2a+\frac{a^{2}}{2})\e^{a}\left(\Gamma(0,a)-\Gamma(0,c)\right)-\frac{3+a}{2}-\e^{-d}\frac{a^{2}(1-c)-4ac}{2c^{2}}\right]\nonumber \\
 &  & +\left(-\frac{A_{lp}}{2}-\frac{D_{l}}{\vec{p}_{T}^{2}}+\frac{E_{lp}}{2\vec{p}_{T}^{2}}\right)\times\nonumber \\
 &  & \left[-\tilde{b}\e^{a}(1+\frac{a}{2})(\Gamma(0,a)-\Gamma(0,c))+\frac{\tilde{b}}{2}\frac{1+a}{a}-\frac{\tilde{b}}{2c^{2}}\e^{-d}(2c-a(1-c))\right]\nonumber \\
 &  & +\left(-2(D_{p}-D_{l})-\frac{E_{l}}{\vec{p}_{T}^{2}}\right)\left[\frac{\tilde{b}^{2}}{2}\e^{a}(\Gamma(0,a)-\Gamma(0,c))+\tilde{b}^{2}\frac{1-a}{2a^{2}}-\tilde{b}^{2}\frac{1-c}{2c^{2}}\e^{-d}\right]\nonumber \\
 &  & +\frac{A_{lp}}{2}\left[-\tilde{b}\e^{a}(1+\frac{a}{2})\Gamma(0,c)+\frac{\tilde{b}}{2c^{2}}(2c-a(1-c))\e^{-d}\right]\nonumber \\
 &  & +\left(-\vec{p}_{T}^{2}\frac{A_{lp}}{2}+D_{l}+\frac{1}{2}E_{lp}\right)\left[\frac{\tilde{b}^{2}}{2}\e^{a}\Gamma(0,c)+\tilde{b}^{2}\frac{1-c}{2c^{2}}\e^{-d}\right]\,.\end{eqnarray}
}{\small \par}
~{}\\

\bibliographystyle{h-physrev3}
\bibliography{Referenzen}

\end{document}